\documentclass[aps,prb,twocolumn,showpacs,amsmath,amssymb,superscriptaddress,floatfix]{revtex4-1}
\usepackage{graphicx}
\usepackage{dcolumn}
\usepackage{bm}
\usepackage{amsfonts}
\usepackage{amsmath}
\usepackage{ulem}
\usepackage{color}
\usepackage{hyperref}

\begin{document}

\title{Photoexcitation of electron wave packets in quantum spin Hall edge states: \\effects of chiral anomaly from a localised electric pulse}

\author{Fabrizio Dolcini}
\email{fabrizio.dolcini@polito.it}
\affiliation{Dipartimento di Scienza Applicata e Tecnologia del Politecnico di Torino, I-10129 Torino, Italy}
\affiliation{CNR-SPIN, Monte S.Angelo - via Cinthia, I-80126 Napoli, Italy}
 
\author{Rita Claudia Iotti}
\affiliation{Dipartimento di Scienza Applicata e Tecnologia del Politecnico di Torino, I-10129 Torino, Italy}
 
\author{Arianna Montorsi}
\affiliation{Dipartimento di Scienza Applicata e Tecnologia del Politecnico di Torino, I-10129 Torino, Italy}
 
\author{Fausto Rossi}
\affiliation{Dipartimento di Scienza Applicata e Tecnologia del Politecnico di Torino, I-10129 Torino, Italy}

\begin{abstract}
We show that, when a spatially localised electric pulse is applied at the edge of a quantum spin Hall system, electron wavepackets of the helical states can be photoexcited by purely intra-branch electrical transitions, without invoking the bulk states or the magnetic Zeeman coupling. 
In particular, as long as the electric pulse remains applied, the photoexcited densities lose their character of right- and left-movers, whereas after the ending of the pulse they propagate in opposite directions without dispersion, i.e. maintaining their space profile unaltered.
Notably we find that, while the momentum distribution of the photoexcited wave packets depends on the temperature~$T$ and the chemical potential $\mu$ of the initial equilibrium state and displays a non-linear behavior on the amplitude of the applied pulse, in the mesoscopic regime  the space profile of the wave packets is independent of $T$ and $\mu$. Instead, it depends purely on the applied electric pulse, in a linear manner, as a signature of the chiral anomaly characterising massless Dirac electrons.  We also discuss how the photoexcited wave packets can be tailored with the electric pulse parameters, for both low and finite frequencies.

\end{abstract}


\maketitle
\section{Introduction}
The realization of an electron-based counterpart of quantum optics, which may yield a dramatic boost in the implementation of quantum information processing, requires the ability to generate, control and detect single electron wave packets.\cite{bertoni2000} 
Recent studies have proposed the exploitation of quantum Hall (QH) systems to this purpose\cite{feve2007,feve2011,bocquillon2013,kataoka2013,bocquillon2014,feve2015,kataoka2015,janssen2016}: semiconductor quantum wells exposed to a perpendicular magnetic field exhibit chiral one-dimensional (1D) edge channels that propagate ballistically and coherently over various micrometers,\cite{roulleau2008} offering an interesting electronic alternative to photonic optical fibers. 
However, large scale applications of QH based electronic devices are limited by the strong values of magnetic fields --various Teslas-- needed to generate the ballistic edge states.

The helical edge states emerging in quantum spin Hall (QSH) effect systems\cite{kane-mele2005a,kane-mele2005b,bernevig_science_2006} may turn the tide in the field. The appearance of these counter propagating states at each edge of narrow gap semiconductor heterostructures, such as HgTe/CdTe\cite{konig_2006,molenkamp-zhang_jpsj,roth_2009,brune_2012} and InAs/GaSb\cite{liu-zhang_2008,knez_2007,knez_2014,spanton_2014} quantum wells, does not require any applied magnetic field, as it originates from a spin-orbit induced topological transition.
Importantly,  helical edge states are also protected from backscattering off non-magnetic impurities, as their  group velocity is locked to their spin orientation. Furthermore, their linear Dirac spectrum implies that a freely propagating electronic wave packet does not undergo the usual dispersion arising in conventional parabolic band materials from the $k$-dependent velocity associated to the various wavepacket components, as  recently shown for the similar case of single walled carbon nanotubes\cite{rosati2015}. 
This is particularly important since the encoding of information in electronic states requires the generation of sequences of wavepackets that propagate  without overlapping, and the control on the wave packet spatial extension and propagation time is crucial to determine the information transmission rate.     
For all these reasons, QSH edge states may be a promising platform for electron quantum optics.\\

\begin{figure} 
\centering
\includegraphics[width=\columnwidth,clip]{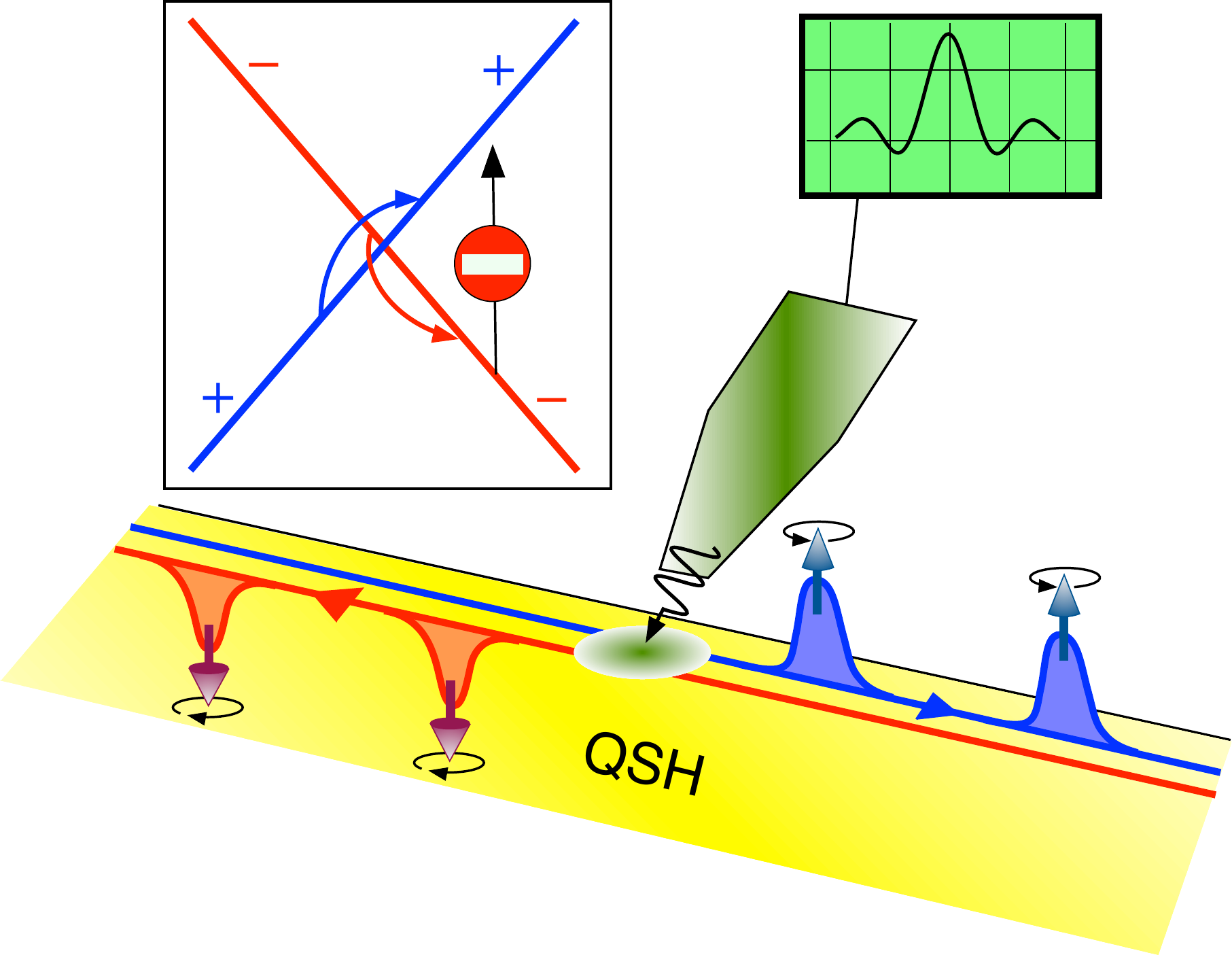}
\caption{(Color online) A spatially localised electric pulse is applied to a quantum spin Hall bar, thereby photoexciting countepropagating electronic wave packets in the helical edge states. The wave packets propagate without dispersion, due to the Dirac linear spectrum of the edge states, with a well defined spin orientation, due to their helical nature. Inset: The two branches $\pm$ of the linear Dirac spectrum of the helical edge states emerging in the topological phase of a QSH system. }
\label{Fig1-setup}
\end{figure}

In order to generate electron wave packets in a controlled way, optical excitation is a widespread strategy. However, important differences emerge between QSH edge states and systems commonly used in optoelectronics. 
In the first instance, the vertical electric dipole transitions that typically occur between valence and conduction bands of conventional semiconductor based devices are forbidden in QSH edge states, due to a selection rule arising from their helical nature (see inset of Fig.\ref{Fig1-setup}).
To circumvent this problem, some works have proposed to exploit circularly polarised radiations, whose magnetic field can induce magnetic dipole transitions on the edge states via Zeeman coupling\cite{cayssol2012,artemenko2013,dolcetto-sassetti2014}. The $g$-factor is, however, rather small. The application of strong magnetic fields has also been considered\cite{kindermann2009}, with the same drawback mentioned for QH edge states, though. 
Alternatively,  for frequencies exceeding the bulk gap, it is possible to induce optical  transitions from the edge states to the bulk states~\cite{artemenko2013,artemenko2015}, which, however, are not topologically protected and exhibit dispersive propagation.  
Most of these approaches are based on the so called far field regime, where a monochromatic radiation is applied over the whole sample for a duration that is long compared to its oscillation period.

There is, however, another reason why  optoelectronics is not trivial in QSH edge states: since they are described by a massless Dirac fermion theory rather than the conventional Schr\"odinger-like parabolic band, their response to an electromagnetic field is intrinsically affected by a peculiar property, known as the chiral anomaly. This subtle behavior, first discovered in the context of the pion decay into photons~\cite{adler1969,bell-jackiw1969}, is on the spotlight in condensed matter physics\cite{burkov2012,li2013,wishvanath2014,takane2016} after the recent observation of 3D Weyl semimetals~\cite{chen2014,hasan2015a,hasan2015b,ong2015}. Since it also affects 1D Dirac fermions, it must be taken into account appropriately in analysing the photoexcitation of wave packets in QSH edge states\cite{trauzettel2016}.\\

In this article, we investigate the response of QSH edge states to an electromagnetic field that is applied on a spatially localised region and for a finite time, as sketched in Fig.\ref{Fig1-setup}. We show that purely intra-branch transitions on the edge states can be induced by an electric pulse directed along the edge, without invoking the bulk states or the Zeeman coupling. As a result, it is possible to photoexcite electron wave packets that are spatially localised and propagate with a well defined spin orientation maintaining their shape without dispersion.  
Notably, we shall show that, despite the momentum distribution of the photoexcited electron wave packets depends on the temperature $T$ and the chemical potential $\mu$ of the initial equilibrium state and exhibits a non-linear dependence on the amplitude $E_0$ of the applied pulse, the space profile of the wave packet is {\it independent} of $T$ and $\mu$. Instead it is purely determined, in a linear manner, by the space- and time- profile of the applied electric pulse. We shall argue that this effect is a signature of the chiral anomaly. \\

The article is organised as follows. After presenting the model and a short account of the chiral anomaly in Sec.\ref{sec-2}, in Sec. \ref{sec-3} we provide some general results concerning the coupling of 1D massless Dirac electrons to a time- and space-dependent electromagnetic field. More specifically, we determine the exact time evolution of the electron field operator for this non-stationary problem and, by exploiting a rigorous procedure that combines regularisation and gauge invariance, we compute the photoexcited electron densities, the space correlations at equal time as well as the time correlations at a local space point, with a particular focus on the momentum distribution and the local tunneling density of states.
Then, in Sec.\ref{sec-4} we focus on the case of a gaussian electric pulse, localized over a finite region $\Delta$ and  characterized by a finite duration~$\tau$ and a frequency~$\Omega$.  We shall explicitly show how the photoexcited wave packets can be tailored with these electric pulse parameters. Finally, in Sec.\ref{sec-5} we discuss our results and hint at possible experimental realisations.
\section{Model}
\label{sec-2} 
The two counter-propagating helical edge states flowing along one boundary of a QSH bar are known to be described by a Dirac massless fermion theory in 1+1 dimensions\cite{molenkamp-zhang_jpsj}, i.e. by a Hamiltonian of the form 
\begin{equation}
\hat{\mathcal{H}}^\circ=v_F \int    dx \,\Psi^\dagger \,    \sigma_z \, \hat{p} \, \Psi^{}\quad. \label{H0}
\end{equation}
In Eq.(\ref{H0})  $v_F$ represents the Fermi velocity playing the role of the speed of light $c$ in high-energy physics,  $x$ and $t$ denote the space and time coordinates, respectively,  while $\hat{p}=-i \hbar \partial_x$ is the momentum operator and $\boldsymbol\sigma=(\sigma_x,\sigma_y, \sigma_z)$ are Pauli matrices acting on the electron two-component spinor field $\Psi(x)$. 
The electronic spectrum of the Hamiltonian (\ref{H0}) consists of two chiral branches with linear dispersion relation, crossing at the Dirac point and describing electrons that propagate rightwards and leftwards along $x$, respectively (see inset of Fig.\ref{Fig1-setup}). The two branches, henceforth labelled by $\pm$, are helical as they also correspond to electrons characterized by two different eigenvalues $\pm 1$ of $\sigma_z$, i.e. to two orthogonal electron fields $\Psi_\pm=P_\pm \Psi=\chi_\pm \psi_\pm$, where $P_\pm=(\sigma_0\pm \sigma_z)/2$ is the $z$-projector, $\sigma_0$ the $2 \times 2$ identity matrix, $\chi_{+}=(1,0)^T$ and $\chi_{-}=(0,1)^T$ the eigenvectors of $\sigma_z$, and $\psi_\pm(x)$ two scalar fields. \\

We shall assume that at the time $t \rightarrow -\infty$  the QSH system is in an equilibrium state, determined by Eq.(\ref{H0}) and characterised by a temperature $T$ and a chemical potential $\mu$. Then, a space- and time-dependent electric pulse $E(x,t)$ is applied along the edge direction $x$, exciting electron density and current.
The Hamiltonian~(\ref{H0}) thus~changes~to
\begin{eqnarray}
\hat{\mathcal{H}}&=&\int    dx \,\Psi^\dagger   \left(v_F  \sigma_z \,(\hat{p}-\frac{\rm e}{c} A(x,t))  + {\rm e} V(x,t) \sigma_0 \right)\Psi^{} = \nonumber\\
&=& \hat{\mathcal{H}}^\circ+ {\rm e}\int dx \, (\hat{n}V-\frac{1}{c}\,\hat{j} A)\quad, \label{H}
\end{eqnarray}
where ${\rm e}$ is the electron charge, and $V$ and $A$ are the scalar and vector potentials yielding the electric field $E(x,t)=-\partial_x V- \frac{1}{c}\partial_t A$, respectively. The Zeeman coupling associated to the magnetic field related to the time variation of $E$ is assumed to be negligible. Note that we deliberately keep the gauge $(V,A)$ in Eq.(\ref{H}) generic. This will enable us to explicitly discuss the gauge independence of our results for the photoexcited density and distribution.
In the second line of Eq.(\ref{H}) $\hat{n}= \Psi^\dagger \Psi^{}=   \hat{n}_{+}+\hat{n}_{-}$ and $\hat{j}= v_F \Psi^\dagger \sigma_z \Psi^{}= v_F (\hat{n}_{+}-\hat{n}_{-})$  are the electron and current densities, respectively, with $\hat{n}_\pm=\Psi^\dagger_{\pm} \Psi^{}_{\pm}$ denoting the density in each chiral branch.  \\

Notably, the Hamiltonian (\ref{H}) still commutes with $\sigma_z$, yielding the important physical consequence mentioned in the introduction: in striking contrast to the case of conventional semiconductors, no vertical optical transition between the lower and the upper Dirac cone is allowed (see inset of Fig.\ref{Fig1-setup}), for this would correspond to a switch in the eigenvalue of $\sigma_z$, which is forbidden by the symmetry of Eq.(\ref{H}). The electric pulse is thus a purely `forward scattering term', in that it does not couple the two $\sigma_z$ helical branches $\Psi_\pm$. It can only induce separate intra-branch transitions $(+) \rightarrow (+)$ and $(-) \rightarrow (-)$, and charge and current are essentially given by the sum and difference of the two dynamically decoupled quantities $\hat{n}_{+}$ and $\hat{n}_{-}$.

Despite such decoupling, determining the photoexcited currents is a non-trivial problem. In the first instance, it is an intrinsically out of equilibrium problem, where the {\it non-stationary} current depends on both the time-dependence and the space profile of the applied pulse $E(x,t)$. 
Secondly, as observed in the introduction, the response of the massless Dirac Hamiltonian (\ref{H0}) to an electromagnetic field is characterised by the chiral anomaly\cite{adler1969,bell-jackiw1969,bertlmann}.  
Because this effect plays a central role in the present paper, before illustrating our results about the photoexcited electronic wave packets, we shall shortly recall the essential aspects of the chiral anomaly, focussing on the case of 1D massless Dirac fermions that is envisaged here.

\subsection{The chiral anomaly in 1+1 dimensions}
To illustrate the chiral anomaly for 1D massless Dirac fermions, we observe that the time evolution of the electron field operator, obtained from the Heisenberg equation of motion dictated by the Hamiltonian (\ref{H}), reads   
\begin{equation}\label{eom-Psi}
i\hbar \partial_t  \Psi  =\left( v_F \sigma_z \, (\hat{p} -\frac{\rm e}{c} A(x,t)) +{\rm e}V(x,t) \sigma_0 \right)\, \Psi \quad,
\end{equation}
and is the massless Dirac equation.\cite{note-Dirac} Combining Eq.(\ref{eom-Psi}) with the equation for $\Psi^\dagger$, one can in principle determine  the dynamical evolution of the field bilinear, such as the electron densities $\hat{n}_\pm=\Psi^\dagger  \Psi^{} =\hat{n}_{+}+\hat{n}_{-}$ or the so called axial density  $\hat{n}^a=\Psi^\dagger \sigma_z \Psi^{}=\hat{n}_{+}-\hat{n}_{-}$. A naive approach to this computation would lead to conclude that two conservation laws  exist  
\begin{eqnarray}
\partial_t \hat{n}(x,t)+\partial_x \hat{j}(x,t)=0  \,\, \, \label{cont-eq}  \\
\partial_t \hat{n}^a(x,t)+  \partial_x \hat{j}^a(x,t)=0 \,\, \, .\label{cont-eq-axial}
\end{eqnarray}
Equation~(\ref{cont-eq}) represents the continuity equation for the electron density, encoding the conservation of the total electron number $\hat{N}=\int \hat{n} \, dx$. Furthermore, Eq.(\ref{cont-eq-axial}) --which only holds for the present case of massless fermions-- is the continuity equation for the axial density $\hat{n}^a$ and axial current $\hat{j}^a=v_F \Psi^\dagger \Psi^{}=v_F(\hat{n}_{+}+\hat{n}_{-})$, which encodes the conservation of the total axial number $\hat{N}^a=\int \hat{n}^a dx$. The existence of these two conserved quantities is seemingly consistent with two symmetries characterizing the equation of motion (\ref{eom-Psi}). Indeed, if $\Psi$ is a solution of Eq.(\ref{eom-Psi}) within the gauge $(V,A)$, then both a {\it gauge} transformation  
\begin{equation}\label{gauge-inv}
\left\{ 
\begin{array}{lc} \Psi(y,t)  \rightarrow   \displaystyle  \Psi^\prime(y,t) = e^{i \chi(x,t)} \Psi(x,t) &
\\  
\begin{array}{lcl}
 \displaystyle V & \rightarrow & \displaystyle V^\prime =V-(\hbar/{\rm e}) \, \partial_t \chi   \\  \displaystyle A & \rightarrow & \displaystyle A^\prime =A+(\hbar c/{\rm e}) \, \partial_x \chi  
\end{array}
&  \end{array}
\right.
\end{equation}
and a {\it chiral} transformation 
\begin{equation}\label{chiral-inv}
\left\{ 
\begin{array}{l} \Psi(x,t)  \rightarrow   \displaystyle  \Psi^\prime(x,t) = e^{i \chi(x,t) \sigma_z} \Psi(x,t)     
\\     
\begin{array}{lcl}
 \displaystyle V & \rightarrow & \displaystyle V^\prime =V-(\hbar/{\rm e} )\partial_t \chi \, \sigma_z \\  \displaystyle A & \rightarrow & \displaystyle A^\prime =A+(\hbar c/{\rm e})\, \partial_x  \chi \, \sigma_z  
\end{array}
\end{array}
\right.
\end{equation}
yield a solution $\Psi^\prime$ of Eq.(\ref{eom-Psi}) within a new gauge $(V^\prime,A^\prime)$ that describes the {\it same} electric field $E(x,t)$ as the original gauge $(V,A)$. Then, the conservation of $\hat{N}$ and $\hat{N}^a$ seems to straightforwardly follow from N\"other's theorem.\\
Importantly, taking sum and difference  of Eqs.(\ref{cont-eq})-(\ref{cont-eq-axial})   would return two equivalent conservation laws
\begin{eqnarray}
\left(\partial_t + v_F  \partial_x \right)\hat{n}_{+}(x,t)=0  \label{cont-right}  \, \,  \\
\left(\partial_t - v_F  \partial_x \right)\hat{n}_{-}(x,t)=0  \label{cont-left}\, \, ,
\end{eqnarray}
i.e. the continuity  equations for each chiral component, which encode the conservation of $\hat{N}_{+}=\int dx \,\hat{n}_{+}$ and $\hat{N}_{-}=\int dx \,\hat{n}_{-}$, separately. 

The seemingly straightforward derivation of these two conservation laws is, however, physically wrong, as can be realised by considering the initial equilibrium state, characterised by a vanishing net current, i.e. a perfect balance between right- and left-moving electrons, $\hat{N}_{+}-\hat{N}_{-}=0$. Then, the separate conservation of $\hat{N}_{+}$ and $\hat{N}_{-}$ would imply that no unbalance $\hat{N}_{+}-\hat{N}_{-} \neq 0$, i.e. no current,  can be induced even when $(V,A) \neq 0$: The metallic electronic system (\ref{H0}) would turn to be inert to any applied electric field,   an obviously unphysical conclusion. 
A physically correct result is thus expected to break the chiral conservation laws (\ref{cont-right}) and (\ref{cont-left}). Equivalently, since the charge conservation (\ref{cont-eq}) must be preserved, the axial conservation law (\ref{cont-eq-axial}) should break down, and for this reason the effect is sometimes referred to as `axial anomaly' as well.\\

The critical point in the above derivation is well known in relativistic quantum electrodynamics, and boils down to the fact that, although Eqs.(\ref{eom-Psi}), (\ref{gauge-inv}) and (\ref{chiral-inv}) are correct, the definition of densities  requires some care\cite{bertlmann}. 
Indeed the Dirac Sea, i.e. the initial equilibrium ground state of (\ref{H0}), contains an infinite number of occupied levels, causing a  divergence in the expectation values of the  bilinear  combinations $\hat{n}_{\pm}=\Psi^\dagger_\pm(x,t) \Psi^{}_\pm(x,t)$  of the fields  evaluated at the same space-time point. 
Note that the presence of divergences is a general feature characterising massless Dirac fermion models: in the Luttinger liquid theory, for instance, where Eq.(\ref{H0}) corresponds to a linearised low energy electronic band, the effects of electron-electron interaction are typically treated by introducing an ultraviolet cutoff and by subtracting the contribution due to the ground state in a controlled way\cite{vondelft}. In the presence of an electromagnetic field, however, this is not sufficient. The mere introduction of a cutoff would lead to results that, despite being finite, depend on the gauge $(V,A)$ chosen for the electromagnetic potentials in (\ref{H}) and violate electric charge conservation.

The physically correct photoexcited wave packet density and current must necessarily be independent of the gauge, obey the charge continuity equation (\ref{cont-eq}) and violate the axial conservation law (\ref{cont-eq-axial}). 
In the next Section, we shall take these aspects into account by combining an exact solution of the electron field operator with the techniques of gauge invariant regularisation to obtain the photoexcited currents.

\section{General results}
\label{sec-3}
In this section we derive some general results concerning the response of 1D massless Dirac electrons to an electromagnetic excitation.
\subsection{Solution of the electron field equation of motion}
We start by proving that the solution of Eq.(\ref{eom-Psi}) is $\Psi(x,t)=\chi_{+} \psi_{+}(x,t)+\chi_{-} \psi_{-}(x,t)$, where  
\begin{equation}\label{Psi-sol-gen}
\psi_{\pm}(x,t)=  \psi^\circ_{\pm}(x \mp v_F t) \,  e^{\pm i \phi_{\pm}(x,t)} \,  \quad.
\end{equation}
In Eq.(\ref{Psi-sol-gen}) the fields $\psi^\circ_{\pm}(x \mp v_F t)$ denote   the space-time evolution of the electron chiral components of Eq.(\ref{H0}), i.e. in absence of the electromagnetic coupling, and describe right-/left-moving electrons, respectively. In contrast, the phases  $\phi_{\pm}$   encode  the effect of the electromagnetic  potentials $V(x,t)$ and $A(x,t)$, and are given by
\begin{eqnarray} 
\phi_{+}(x,t)    & =  &   \frac{{\rm e} v_F}{\hbar c} \int_{-\infty}^t \!\! (A-\frac{c}{v_F}V)(x-v_F (t-t^\prime), t^\prime) \, dt^\prime = \nonumber \, \,  \\ 
&= & \frac{{\rm e}}{\hbar c} \int_{-\infty}^x \, (A-\frac{c}{v_F}V)(x^\prime, t-\frac{x-x^\prime}{v_F}) \, dx^\prime  \label{phi+} 
\end{eqnarray}
and
\begin{eqnarray}
\phi_{-}(x,t)    &= &  \displaystyle  \frac{{\rm e} v_F}{\hbar c} \int_{-\infty}^t \!\! (A+\frac{c}{v_F}V)(x+v_F (t-t^\prime), t^\prime) \, dt^\prime = \nonumber  \\ 
&= & \displaystyle  \frac{{\rm e}}{\hbar c} \int_{x}^\infty \, (A+\frac{c}{v_F}V)(x^\prime, t+\frac{x-x^\prime}{v_F}) \, dx^\prime \quad. \label{phi-}
\end{eqnarray}
The above expressions have a straightforward physical interpretation. The first (second) line of Eq.(\ref{phi+}), for instance,   expresses the phase $\phi_+(x,t)$ induced by the electromagnetic field on the right-moving electron $\psi^\circ_{+}(x-v_F t)$ as a convolution over time (space) of the values of the electromagnetic potentials at times earlier than~$t$ and at positions located on the left of~$x$,   propagating with the electron Fermi velocity $v_F$ according to the   dynamics dictated by Eq.(\ref{H0}). A similar  result is expressed for the phase $\phi_{-}(x,t)$ in Eq.(\ref{phi-}). Notice that, while in high energy Physics the propagation of massless Dirac fermions occurs at the same speed as the electromagnetic wave, $c$, here it is characterised by the Fermi velocity $v_F$. We emphasise that, under the only constraint that the electromagnetic potentials vanish for $t \rightarrow -\infty$, the solution (\ref{Psi-sol-gen}) is valid for arbitrary $V(x,t)$ and $A(x,t)$, so that the space- and time-dependence of the induced phases $\phi_\pm(x,t)$ is not necessarily of the form $\phi_\pm(x \mp v_F t)$. As a consequence, in the presence of the electromagnetic pulse the electron fields (\ref{Psi-sol-gen}) lose their right- and left-moving character, despite the absence of any `back-scattering' that couples them. \\

The proof of Eq.(\ref{Psi-sol-gen}) follows from noticing that, by decomposing $\Psi(x,t)=\chi_{+} \psi_{+}(x,t)+\chi_{-} \psi_{-}(x,t)$ in  the two scalar chiral components $\psi_\pm$,  Eq.(\ref{eom-Psi}) is equivalent  to a set of two decoupled equations, 
\begin{equation}\label{eq-L}
\hat{L}_\pm \,   \psi_\pm  =  \frac{{\rm e} v_F}{c} \left(\frac{c}{v_F} V(x,t) \mp A(x,t) \right)   \psi_\pm  \quad,
\end{equation} 
where 
$
\hat{L}_\pm=i \hbar\, (\partial_t  \pm v_F \partial_x) 
$
are  chiral operators characterising the equation of motion $\hat{L}_\pm\, \psi^\circ_\pm =0$ dictated by the free Hamiltonian Eq.(\ref{H0}).
Exploiting the retarded Green functions $G^{\rm ret}_\pm$ of such operators,  
$\hat{L}_\pm \, G^{\rm ret}_\pm(x,t;x^\prime, t^\prime)=\delta(x-x^\prime) \, \delta (t-t^\prime)$, it is straightforward to see that the expression 
\begin{equation}\label{psipm-sol-method1-pre}
\psi_\pm (x,t)= e^{\frac{{\rm e} v_F}{c} \int d\mathbf{x}^\prime G^{\rm ret}_\pm(\mathbf{x},\mathbf{x}^\prime) (\frac{c}{v_F}V\mp A)(\mathbf{x}^\prime) } \, \psi^\circ_\pm (x \mp v_F t)\quad,
\end{equation}
with $\mathbf{x}=(x,t)$ and $\mathbf{x}^\prime=(x^\prime,t^\prime)$, solves Eq.(\ref{eq-L}). By substituting into Eq.(\ref{psipm-sol-method1-pre}) the explicit expression $i \hbar G^{\rm ret}_\pm(x,t;x^\prime, t^\prime)=\, \theta(t-t^\prime) \, \delta(v_F(t-t^\prime)\mp (x-x^\prime))$ for the retarded Green function, one finds the solution (\ref{Psi-sol-gen})  with Eqs.(\ref{phi+}) and (\ref{phi-}).\\

The obtained time evolution (\ref{Psi-sol-gen}) of the electron field operator enables us to evaluate the expectation values of electron fields, such as bilinears $\langle \psi^\dagger_\pm (x_1,t_1) \psi^{}_\pm (x_2,t_2) \rangle_\circ$ or higher order correlations. Because we adopt the Heisenberg picture, the whole time dependence is attributed here to the fields (\ref{Psi-sol-gen}), with $\langle \ldots \rangle_\circ={\rm Tr}[\ldots \hat{\rho}^\circ_\pm]$ denoting the average value with respect to the time-independent equilibrium density matrix $\hat{\rho}^\circ={\rm diag}(\hat{\rho}^\circ_{+},\hat{\rho}^\circ_{-})$ at $t = -\infty$ stemming from the Hamiltonian~(\ref{H0}) and characterised by a temperature $T$ and a chemical potential $\mu$. In next subsection we shall provide general results about photoexcited densities and current as well as electronic correlations.


\subsection{Space profile of photoexcited electron and current densities}
\label{sec-3B}
The photoexcited electron and current densities, defined as $\Delta n=\langle \hat{n}-\hat{n}^\circ\rangle_\circ$ and $\Delta j=\langle \hat{j}-\hat{j}^\circ\rangle_\circ$, respectively, identify the deviations in the expectation value of electron and current densities induced by the electromagnetic field with respect to the initial equilibrium values $\hat{n}^\circ$ and $\hat{j}^\circ$. They can be straightforwardly expressed through  
\begin{eqnarray}\label{photo-dens}
\Delta n &= &\Delta n_{+} \, + \, \Delta n_{+} \label{photo-dens} \\
\Delta\hat{j} &= &v_F (\Delta n_{+} \, - \, \Delta n_{+}) \label{photo-curr}
\end{eqnarray}
in terms of the photoexcited chiral densities $\Delta n_{\pm}=\langle \hat{n}_{\pm}-\hat{n}^\circ_{\pm}\rangle_\circ =\langle \psi^\dagger_\pm \psi^{}_\pm  -{\psi^\circ}^\dagger_\pm \psi^{\circ}_\pm \rangle_\circ$.
Since the electromagnetic coupling merely affects the phase of the electron field [see Eq.(\ref{Psi-sol-gen})], one would be tempted to conclude that $\hat{n}_\pm(x,t)=\psi^\dagger_\pm \psi^{}_\pm={\psi^\circ_\pm}^\dagger \psi^{\circ}_\pm = \hat{n}^\circ_\pm(x \mp v_F t)$, i.e. that the densities remain unaffected ($\Delta n_{\pm}=0$), thereby recovering the separate conservation of $\hat{N}_{+}$ and $\hat{N}_{-}$ and Eqs.(\ref{cont-right})-(\ref{cont-left}). As observed above, this conclusion is wrong and an account of  the infinite number of occupied states  characterising the initial equilibrium state is mandatory.

\subsubsection{Regularisation with gauge invariance}
Here we describe the technical procedure to obtain physically correct results for photoexcitations in massless Dirac fermions. Two physical principles underlie the definition of suitable operators $\Delta \hat{n}_\pm$ that determine the photoexcited density and current (\ref{photo-dens}) and (\ref{photo-curr}): i) finite measurable quantities can only be extracted upon controlling the divergence due to the equilibrium Dirac Sea; ii) the result must be independent of the specific gauge chosen to describe the electric field $E(x,t)$. Note that, just like the Hamiltonian (\ref{H}), the dynamical evolution (\ref{Psi-sol-gen}) of the electron field operator does depend on the gauge, as appears from inspection of Eqs.(\ref{phi+})-(\ref{phi-}).  To fulfil the two  above requirements, one defines\cite{bertlmann}
\begin{eqnarray}
 \Delta n_{\pm}(x,t)  \doteq  
 \!\! \lim_{(x^\prime,t^\prime) \rightarrow (x,t) } \left\langle  \psi^\dagger_{\pm}(x^\prime,t^\prime)\psi^{}_{\pm}(x,t)\, e^{-iW_L(x,t,x^\prime,t^\prime)}  \, \right.      \nonumber \\\nonumber  \\  \left. - {\psi^\circ_{\pm}}^\dagger(x^\prime,t^\prime){\psi^\circ_{\pm}}^{}(x,t)  \right\rangle_\circ \hspace{1cm}  \label{Deltan_{pm}-def-gen} 
\end{eqnarray}
where $\psi_{\pm}(x,t)$ are the electron field operators in the presence of the electromagnetic field,  $\psi^\circ_{\pm}(x,t)$ are the ones in  absence of electromagnetic field, and
\begin{equation}
W_L(x,t,x^\prime,t^\prime)=\frac{\rm e}{\hbar c} \int_{(x,t)}^{(x^\prime,t^\prime)} (c V d  t^{\prime\prime}-A \, dx^{\prime\prime}) \label{WL-def}
\end{equation}
is a Wilson line, i.e. a contour integral of the electromagnetic potentials $(V,A)$, performed in the space-time along any path connecting the two split points $(x^\prime,t^\prime)$ and $(x,t)$.

The two physical principles mentioned above are implemented in the definition  (\ref{Deltan_{pm}-def-gen}) by two mathematical ingredients.
The first one is the point-splitting, i.e. the fact that the field bilinear is evaluated as the limit for two {\it different} arguments $(x^\prime,t^\prime)\neq (x,t)$ in the fusing fields. To avoid any spurious dependence on the limit direction the standard procedure is to introduce an infinitesimal vector ${\epsilon^\mu}=(v_F \epsilon_t,\epsilon_x)=(v_F(t^\prime-t),x^\prime-x)$ in the space-time and to perform the limit according to the Minkowski metric tensor $\eta^{\mu \nu}$, i.e. 
$
\lim_{\epsilon \rightarrow 0}  \epsilon^\mu \epsilon^\nu/\epsilon^2 = \eta^{\mu \nu} 
$.
The point splitting enables one to handle  the diverging contribution of the ground state.   
However, in applying it,  the gauge invariance of the bilinear   combination is explicitly broken by an amount $\chi(x,t)-\chi(x^\prime,t^\prime)$, where $\chi$ is the function identifying a gauge transformation [see Eq.(\ref{gauge-inv})]. Although small when $(x^\prime,t^\prime) \rightarrow (x,t)$, such gauge dependent amount does yield a finite contribution to $\Delta \hat{n}_{\pm}$ when combined with the diverging expectation value $\langle {\psi^\circ_{\pm}}^\dagger(x^\prime,t^\prime){\psi^\circ_{\pm}}^{}(x,t) \rangle_\circ$.
Thus, in order to obtain gauge invariant $\Delta \hat{n}_\pm$, the introduction of the second ingredient is needed, namely the phase associated to the Wilson line (\ref{WL-def}), which   compensates for the gauge phase difference $\chi(x,t)-\chi(x^\prime,t^\prime)$ acquired by the fields upon point-splitting. 

The role of the Wilson line can be appreciated by making a comparison with the case of conventional Schr\"odinger-like model characterised by a parabolic band with an effective mass $m^*$, and by considering the current operator, given in that case by $\hat{j}=({\rm e}/m^*) [ \Psi^\dagger (\hat{p}-\frac{\rm e}{c}A)\, \Psi +{\rm H.c.}]$. The  gauge-dependent term $\frac{\rm e}{c}A$ compensates for the gauge-dependence arising from the non local action of the operator $\hat{p}=-i \hbar \partial_x$, in order to give a gauge invariant expectation value $\langle \hat{j} \rangle$. In contrast, in the massless Dirac model, the current operator $\hat{j}=v_F \Psi^\dagger \sigma_z \Psi$ is local in space and, as a consequence, does not carry any explicit dependence on the vector potential $A$. However, as observed above, the non-locality is subtly hidden in the field point splitting that is needed to deal with the divergent contribution of the equilibrium ground state. The Wilson line thus plays the same role  as the term $\frac{\rm e}{c}A$ in a Schr\"odinger-like model in  restoring the gauge invariance. 

It is also worth emphasising that, despite its mathematical aspect, the  procedure (\ref{Deltan_{pm}-def-gen}) of point-splitting equipped with the Wilson line is not a merely formal issue. Indeed any numerical implementation of massless Dirac fermions  requires the introduction of an ultraviolet cut-off $k_{max}$, which in fact corresponds to performing a point-splitting $\psi^\dagger_{\pm}(x\pm i \frac{a}{2},t)\psi^{}_{\pm}(x\mp  i \frac{a}{2},t)$ where the space coordinates are separated by a small imaginary part $a=1/k_{max}$,  thereby  breaking gauge invariance. 
Thus, without the Wilson line (\ref{WL-def}) in Eq.(\ref{Deltan_{pm}-def-gen}), one would obtain finite results for $\Delta n_\pm$ and for the density and current (\ref{photo-dens})-(\ref{photo-curr}), which, however, would be gauge dependent and violate charge conservation.

\subsubsection{Explicit expressions} 
Applying the regularisation procedure described above, we have computed the photoexcited helical densities $\Delta n_{\pm}$. The result can be given four equivalently useful  expressions, namely
\begin{eqnarray}
\Delta n_{+}(x,t)      
&=& \displaystyle +\frac{\rm e}{2\pi\hbar}   \int_{-\infty}^{t}   \, E(x-v_F (t-t^\prime), t^\prime)  \,   dt^\prime = \nonumber\\  
&=& \displaystyle+\frac{{\rm e}}{2\pi\hbar  v_F}   \int_{-\infty}^{x}   \, E(x^\prime, t-\frac{x-x^\prime}{v_F})  \,   dx^\prime=  \nonumber \\
&=& \displaystyle  + \frac{1}{2\pi}  \left( \partial_x \phi_{+}(x,t) - \frac{\rm e}{\hbar c} A(x,t) \right)   =  \nonumber \\
&=& \displaystyle  - \frac{1}{2\pi v_F}  \left(  \partial_t \phi_{+}(x,t)+  \frac{\rm e}{\hbar } V(x,t) \right)   \label{Deltan_{+}_res-gen}
\end{eqnarray}
and
\begin{eqnarray}
\Delta n_{-}(x,t)  &=&  \displaystyle-\frac{\rm e}{2\pi\hbar}   \int_{-\infty}^{t}   \, E(x+v_F (t-t^\prime), t^\prime)  \,   dt^\prime \, = \nonumber\\ 
&=& \displaystyle -\frac{{\rm e}}{2\pi\hbar  v_F}   \int_{x}^{\infty}   \, E(x^\prime, t+\frac{x-x^\prime}{v_F})  \,   dx^\prime = \nonumber\\
&=& \displaystyle  + \frac{1}{2\pi}  \left( \partial_x \phi_{-}(x,t) + \frac{\rm e}{\hbar c} A(x,t) \right)   =\nonumber  \\
&=& \displaystyle  + \frac{1}{2\pi v_F}  \left(  \partial_t \phi_{-}(x,t)-  \frac{\rm e}{\hbar } V(x,t) \right) 
\label{Deltan_{-}_res-gen}
\end{eqnarray}

The expressions in the third and fourth lines of Eqs.(\ref{Deltan_{+}_res-gen})-(\ref{Deltan_{-}_res-gen}) have been obtained by inserting the electron field evolution (\ref{Psi-sol-gen}) into Eq.(\ref{Deltan_{pm}-def-gen}),  
\begin{eqnarray}
\Delta n_{\pm}(x,t)  \doteq  
 \!\! \lim_{(x^\prime,t^\prime) \rightarrow (x,t) }  \left\{ \left\langle {\psi^\circ_{\pm}}^\dagger(x^\prime,t^\prime){\psi^\circ_{\pm}}^{}(x,t)  \right\rangle_\circ  \times \right.  \hspace{1cm}     \nonumber \\\nonumber  \\  \hspace{1cm} \times \left. \left( e^{i(\pm (\phi_\pm(x,t)-\phi_\pm(x^\prime,t^\prime)) -W_L(x,t,x^\prime,t^\prime))} -1 \right) \right\}\, ,  \hspace{1cm}  \label{Deltan_{pm}-res-gen} 
\end{eqnarray}
and by performing the limit as discussed above, exploiting the field-free correlation 
\begin{eqnarray}
\lefteqn{\langle {\psi^\circ_{\pm}}^{\dagger}(x^\prime,t^\prime) {\psi^\circ_{\pm}}^{}(x,t)  \rangle_\circ =  \hspace{3cm} } & & \nonumber \\
& & =   \frac{\pm  i\, e^{\pm i k_F \left( x^\prime-x \mp v_F (t^\prime-t)\right)}}{2 l_T \sinh\left[  \left( x^\prime-x \mp v_F (t^\prime-t)\right) \frac{\pi}{l_T}\right]} \quad,\label{correlations0} 
\end{eqnarray}
with $l_T=\beta \hbar v_F=\hbar v_F/k_B T$ denoting the thermal length and $k_F=\mu/\hbar v_F$ the Fermi wavevector of the initial equilibrium state. The expressions in the first and second line  of Eq.(\ref{Deltan_{+}_res-gen}) [Eq.(\ref{Deltan_{-}_res-gen})] are then obtained by substituting Eq.(\ref{phi+}) [Eq.(\ref{phi-})] into either the third or the fourth line, and by exploiting the hypothesis that $V$ and $A$ vanish for $t \rightarrow -\infty$.\\

\noindent The properties of the obtained results (\ref{Deltan_{+}_res-gen})-(\ref{Deltan_{-}_res-gen}) are noteworthy.\\

i) Eqs.(\ref{Deltan_{+}_res-gen})-(\ref{Deltan_{-}_res-gen}) show that the space profiles $\Delta \hat{n}_{\pm}$ depend {\it only} on the applied electric pulse $E$, in a linear manner, whereas they are {\it independent} of the temperature and chemical potential characterising the initial equilibrium state. This behavior is a peculiarity of the linear spectrum of massless Dirac fermions, and would be absent in the presence of band curvature. At a more formal level, this stems from the conformal invariance of massless Dirac theory, which causes the correlation functions to display the simple scaling laws of a critical system: thus, in the field-free correlation (\ref{correlations0}) the limit $|x^\prime-x|\,, \,|t^\prime-t|\rightarrow 0$ of small space and time difference is equivalent to rescaling the temperature and the Fermi wave vector to zero (i.e. $l_T=\hbar v_F/k_B T \rightarrow \infty$ and $k_F \rightarrow 0$), so that the dependence on these quantities drops out.\\

ii) {\it Chiral anomaly.}  From the first line of the obtained solutions Eqs.(\ref{Deltan_{+}_res-gen})-(\ref{Deltan_{-}_res-gen}) one can prove that  
\begin{equation}  
\begin{split}
(\partial_t + v_F \partial_x)  \!\Delta   \hat{n}_{+}(x,t)   \, \,  =  + \frac{\rm e}{2\pi\hbar}  E(x,t)  \, \,     \\
(\partial_t - v_F \partial_x)  \!\Delta   \hat{n}_{-}(x,t)   \, \,  =  - \frac{\rm e}{2\pi\hbar}  E(x,t)   \, \, ,
\end{split} \label{cont-right-left-new}
\end{equation}
which replace the unphysical conservation laws (\ref{cont-right})-(\ref{cont-left})  of chiral currents by displaying on their right-hand side an anomalous term describing the response to the electric field $E(x,t)$\cite{nielsen1983}.  
Equivalently,  the sum and difference of Eqs.(\ref{cont-right-left-new}) yield 
\begin{eqnarray} 
\partial_t \Delta\hat{n}(x,t)+\partial_x \Delta\hat{j}(x,t)=0 \hspace{2cm}  \\
\partial_t \Delta\hat{n}^a(x,t)+ \partial_x \Delta\hat{j}^a(x,t)=\frac{\rm e}{\pi\hbar}  E(x,t) \quad.
\label{eq-axial-anomaly}
\end{eqnarray}
While the continuity equation is fulfilled, the axial charge is not conserved due to the anomalous term appearing in Eq.(\ref{eq-axial-anomaly}). 
Notably, the anomalous term on the r.h.s. of Eq.(\ref{eq-axial-anomaly}) [or equivalently in Eqs.(\ref{cont-right-left-new})] depends {\it only} on the universal constant ${\rm e}/2\pi \hbar$ and the electric field $E(x,t)$, and not on the electron degrees of freedom. This shows the close relation between the chiral anomaly and the above  mentioned $T$- and $\mu$-independence of $\Delta n_{\pm}(x,t)$ for massless Dirac fermions. Indeed for massive fermions with a non-linear dispersion relation, Eq.(\ref{eq-axial-anomaly})  would display an additional (non-anomalous) term, proportional to the mass and dependent on the electronic state\cite{bertlmann}.  \\

 iii) The equalities in the first and second lines of Eqs.(\ref{Deltan_{+}_res-gen})-(\ref{Deltan_{-}_res-gen}) directly express the photoexcited chiral densities $\Delta \hat{n}_\pm$ in terms of time and space convolution of the electric field $E(x,t)$, thereby proving the gauge-invariance of the result. In contrast, the equalities appearing in the third (fourth) lines of Eqs.(\ref{Deltan_{+}_res-gen})-(\ref{Deltan_{-}_res-gen}) express $\Delta \hat{n}_\pm$ as a combination of the space (time) derivative of the phases $\phi_{\pm}$, given by Eqs.(\ref{phi+})-(\ref{phi-}), and of the vector (scalar) potential. The latter term stems from the Wilson line, and compensates for the gauge dependence of the former $\phi_{\pm}$-term, yielding a gauge-invariant result; \\

iv) From the first line on the right-hand side Eqs.(\ref{Deltan_{+}_res-gen})-(\ref{Deltan_{-}_res-gen}) one can straightforwardly prove that
$
\Delta \hat{N}=\Delta \hat{N}_{+}+\Delta \hat{N}_{-}=0$, with $\Delta \hat{N}_{\pm}=\int_{-\infty}^{+\infty} \Delta \hat{n}_\pm(x,t)\, dx
$, i.e.  the {\it total} charge created by the electromagnetic field is always vanishing, as expected in a photoexcitation problem. Notice, however, that in general $\Delta \hat{N}_{+}=-\Delta \hat{N}_{-} \neq 0$: for massless Dirac electrons the electric pulse does not merely redistribute  electronic states within each helical branch $+$ and $-$; rather, it effectively `creates' electrons in one branch and `depletes' the other branch accordingly. This peculiarity of massless Dirac electrons  can be illustrated by considering  the example of a uniform applied electric field $E_0$, which causes a shift $k \rightarrow k-{\rm e} E_0 t/\hbar$ in the wave vector. Because there is no lower bound in $k$, a comparison with the initial ground state shows that at a time $t$ the net effect of the field corresponds to an effective creation of electrons in one branch, with a corresponding depletion of electrons in the other branch. This phenomenon can be regarded to as an effective transfer of electrons from one branch to the other, via the depth of the Dirac Sea, despite the Hamiltonian (\ref{H}) does not explicitly couple the branches.  \\

\subsection{Equal time correlations: density matrix and momentum distribution}
The equal time  correlations are described by the density matrix $\hat{\rho}$, which in turn enables one to compute the expectation value of any single-particle observable $\hat{O}$ as $ \langle \Psi^\dagger \hat{O}\Psi \rangle_\circ={\rm Tr}[\hat{\rho} \,\hat{O}]$. In its real space representation the density matrix entries $\rho(x_2,x_1;t)= \langle \Psi^\dagger(x_1,t)\Psi^{}(x_2,t)\rangle_\circ$ can be regarded
 as the generalisation of the space density $n(x,t)=\langle \Psi^\dagger(x,t)\Psi^{}(x,t)\rangle_\circ$ to off-diagonal space points. As we discussed in the previous section, the inclusion of a Wilson line is crucial to obtain a physically correct density. Thus, for the case of massless Dirac fermions it is natural to introduce a gauge-invariant density matrix that includes a Wilson line, 
\begin{eqnarray}
\lefteqn{\rho(x_2,x_1;t)   \doteq  e^{+\frac{i{\rm e}}{\hbar c}  \int_{x_2}^{x_1}   A(x^\prime,t)  \, d x^\prime} \times}  & &  \label{rho-mat-def} \\& & \nonumber  \\
& \times &   
\left( \begin{array}{cc} \left\langle {\psi}_{+}^\dagger(x_1,t)   \psi_{+}(x_2,t)   \right\rangle \,  & \left\langle {\psi}_{+}^\dagger(x_1,t)   \psi_{-}(x_2,t)   \right\rangle \\
& \\
 \left\langle {\psi}_{-}^\dagger(x_1,t)   \psi_{+}(x_2,t) \right\rangle  & \left\langle  \psi_{-}^\dagger(x_1,t)  \psi_{-}(x_2,t)   \right\rangle \,  
\end{array} \right)   \,   \nonumber
\end{eqnarray}
Notice that, because the density matrix is an {\it equal time} bilinear, the Wilson line (\ref{WL-def}) reduces to the phase pre-factor in Eq.(\ref{rho-mat-def}) that  only involves the vector potential~$A$. Thus, for a gauge with a  purely {\it scalar} potential $V(x,t)$, Eq.(\ref{rho-mat-def})  coincides with the ordinary density matrix.\\

Because for the present case the  $\sigma_z$ projections are dynamically decoupled in the Hamiltonian (\ref{H}), the density matrix is block-diagonal in the spinor basis, $\hat{\rho}={\rm diag}(\hat{\rho}_{+},\hat{\rho}_{-})$, where the blocks in the real space representation are explicitly obtained by the solution Eq.(\ref{Psi-sol-gen}),
\begin{eqnarray}
\rho_{\pm}(x_2,x_1;t) & =& \langle \psi^\dagger_{\pm}(x_1,t)\psi^{}_{\pm}(x_2,t)\rangle_\circ e^{ \frac{i{\rm e}}{\hbar c}\int_{x_2}^{x_1} \!\!A(x^\prime, t) \, dx^\prime }= \nonumber    \\
&=&  \langle {\psi^\circ_{\pm}}^\dagger(x_1,t) {\psi^\circ_{\pm}}^{}(x_2,t)  \rangle_\circ  \times \nonumber \\
& & \times \, e^{\mp i  (\phi_{\pm}(x_1,t) -\phi_{\pm}(x_2,t) \mp  \frac{{\rm e}}{\hbar c}\int_{x_2}^{x_1} \!\!A(x^\prime, t) \, dx^\prime) } \hspace{0.5cm}\label{rho_{pm}-def}
\end{eqnarray}
where $\langle {\psi^\circ_{\pm}}^\dagger(x_1,t) {\psi^\circ_{\pm}}^{}(x_2,t) \rangle_\circ$ is given by Eq.(\ref{correlations0}) and $\phi_{\pm}$ by Eqs.(\ref{phi+})-(\ref{phi-}). 
In particular, the photoexcited density matrix, describing the {\it deviations} $\Delta \rho_{\pm}$ induced by the electromagnetic field on the equilibrium density matrix $\rho^\circ_{\pm}$, is given by
\begin{eqnarray}
\lefteqn{\Delta\rho_{\pm} (x-\frac{y}{2},x+\frac{y}{2};t)  \doteq  } & & \nonumber \\
&=& \rho_{\pm} (x-\frac{y}{2},x+\frac{y}{2},t) \, - \,  \rho^\circ_{\pm} (x-\frac{y}{2},x+\frac{y}{2},t)
= \label{Deltarho{pm}-res-pre} \\
&  & =\left\langle {\psi^\circ_{\pm}}^\dagger(x+\frac{y}{2},t) {\psi^\circ_{\pm}}^{}(x-\frac{y}{2},t) \right\rangle_\circ \, \,\left( e^{\mp i  \Delta \phi^{\rm et}_{\pm}(x,y;t) }  \, -1\right) 
\nonumber
\end{eqnarray}
where we have introduced the `center-of-mass' $x =(x_1+x_2)/2$ and the relative coordinate $y = x_1-x_2$, and the  equal time gauge invariant phase difference
\begin{eqnarray}
\lefteqn{\Delta \phi^{\rm et}_{\pm}(x,y;t) \doteq } & & \label{Deltaphig-et-def} \\
& & \phi_{\pm}(x+\frac{y}{2},t) -\phi_{\pm}(x-\frac{y}{2},t) \mp  \frac{{\rm e}}{\hbar c}\int_{x-\frac{y}{2}}^{x+\frac{y}{2}} \!\!\!A(x^\prime, t) \, dx^\prime ,\nonumber
\end{eqnarray}
with $\phi_\pm$ given by Eqs.(\ref{phi+})-(\ref{phi-}). 
With the use of Eq.(\ref{correlations0}), it can easily be checked that,  in the diagonal limit $y  \rightarrow 0$,  one recovers from Eq.(\ref{Deltarho{pm}-res-pre})  the gauge invariant chiral densities $\Delta n(x,t)$ [see third lines of Eqs.(\ref{Deltan_{+}_res-gen})-(\ref{Deltan_{-}_res-gen})].\\

The momentum space representation of the gauge-invariant density matrix can straightforwardly be obtained by Fourier transform
\begin{eqnarray}\label{rho_{pm}-k-def}
\rho_{\pm}(k_2,k_1;t)  = \frac{1}{L} \iint dx_1    dx_2 \, e^{i (k_1 x_1- k_2 x_2)} \rho_{\pm}(x_2,x_1;t) \hspace{0.7cm}
\end{eqnarray}
where $L$ denotes the length of the 1D edge states systems, and is assumed to be the longest length scale in the problem ($L  \rightarrow \infty$). By substituting Eq.(\ref{rho_{pm}-def}) into Eq.(\ref{rho_{pm}-k-def}), two equivalent expressions can be given for the result. The first one, 
\begin{eqnarray}
 \rho_\pm(k_2, k_1;t) 
= L \!\int \frac{d k^\prime}{2\pi}  \,  \Gamma_{k_1-k^\prime}^{\pm}(t)   \Gamma_{k_2-k^\prime}^{\pm  {}^{*}}(t)    f^\circ_\pm(k^\prime) \, e^{\pm k^\prime a}   , \hspace{0.6cm} \label{rho-k-kprime-convk}
\end{eqnarray}
expresses the entries of the density matrix in terms of the equilibrium Fermi distribution $f^\circ_\pm(k) = \left\{ 1+\exp \left[\beta \hbar v_F (\pm k-k_F) \right] \right\}^{-1}$
and a set of dimensionless coefficients  
\begin{equation} 
\Gamma^\pm_{k}(t)  =  \frac{1}{L}\int  e^{i k x}  \, e^{\mp i  \phi^{\rm et}_{\pm}(x,t)   }  dx    \label{Gamma_{pm}-def}
\end{equation}
that encode the effect of the electromagnetic field on each $k$-state, with 
$\phi^{\rm et}_{\pm}(x,t) = \phi_{\pm}(y,t) \mp  ({\rm e}/\hbar c) \int_{0}^{x} A(x^\prime,t) \, dx^\prime $, and $a$ denoting the ultraviolet cut-off length. In particular, the momentum distribution, given by the diagonal entries $k_2=k_1=k$ of Eq.(\ref{rho-k-kprime-convk}),   
\begin{eqnarray}
f_\pm(k;t) 
= L \int \frac{dk^\prime}{2\pi}  \,  |\Gamma_{k-k^\prime}^{\pm}(t)|^2    f^\circ_\pm(k^\prime) \, e^{\pm k^\prime a}   \, \,  \label{f-k-first}
\end{eqnarray}
appears as a convolution of the squared $\Gamma_\pm$-coefficients, induced by the electromagnetic field, weighted by the initial equilibrium distribution.

The second expression for the momentum representation of the  density matrix can be obtained by switching integration variables $(x_1,x_2) \rightarrow (x,y)$ in Eq.(\ref{rho_{pm}-k-def}), and by introducing the  average momentum $k=(k_1+k_2)/2$ and the transferred momentum $q=k_1-k_2$, 
\begin{eqnarray}\label{rho_{pm}-k-def-2}
\rho_{\pm}(k,q;t)  =  \iint \frac{dx  \,  dy}{L} e^{i (q x+ k y)} \rho_{\pm}(x-\frac{y}{2},x+\frac{y}{2};t) \,.\hspace{0.7cm}
\end{eqnarray}
By exploiting  the formula  
\begin{eqnarray}
\pm i \lim_{a \rightarrow 0} \int_{-\infty}^{+\infty} \frac{F(y)}{l_T \, \sinh[\pi (y\pm ia)/l_T]}\, dy \, =  \nonumber \\
= F(0) \, \pm i \,\mathcal{P} \int_{-\infty}^{+\infty} \frac{F(y)}{l_T \, \sinh[\pi y /l_T]}\, dy \, 
\end{eqnarray}
where $F$ is an arbitrary function and $\mathcal{P}$ denotes the principal value, and by using the property that (\ref{Deltaphig-et-def}) is odd in the relative variable $y$, one obtains
\begin{eqnarray}
\rho_{\pm}(k,q;t)  =  \frac{\delta_{q,0}}{2} \,+\frac{1}{2 L} \frac{1}{l_T }   \int_{-\infty}^{+\infty} \!\!\! dx\, e^{i q x}  \, \, \times \nonumber \\
\times \int_{-\infty}^{+\infty}\frac{\sin\left(\Delta\phi^{\rm et}_{\pm}(x,y;t)   \mp (k \mp k_F) y  \right)}{\sinh[\pi y/l_T]}\, dy \quad.   \label{rhok_{pm}-fin}
\end{eqnarray}
Then, the momentum distribution is also obtained  from (\ref{rhok_{pm}-fin}) by taking $q=0$, i.e. $f_{\pm}(k;t)\doteq \rho_{\pm}(k,q=0;t)$.\\

In particular, the momentum distribution  of the photoexcited wave packets reads
\begin{eqnarray}
\lefteqn{\Delta f_{\pm}(k;t)  \doteq f_{\pm}(k;t) -f^\circ_{\pm}(k) = } & & \nonumber \\
&=&  \frac{1}{2 L} \frac{1}{l_T }   \int_{-\infty}^{+\infty} \!\!\! dx\,  \int_{-\infty}^{+\infty}  \, dy \,  \label{Deltafk_{pm}-fin} \\
& & \times \,\frac{\sin\left[\Delta\phi^{\rm et}_{\pm}(x,y;t)   \mp (k \mp k_F) y  \right]\pm \sin\left[ (k \mp k_F) y  \right]}{\sinh[\pi y/l_T]}\quad.   \nonumber 
\end{eqnarray}
Importantly, the photoexcited momentum distribution~(\ref{Deltafk_{pm}-fin}) depends on the temperature $T$ and on the chemical potential $\mu$, in striking contrast with the photoexcited density profiles $\Delta n_{\pm}(x,t)$ [see Eqs.(\ref{Deltan_{+}_res-gen})-(\ref{Deltan_{-}_res-gen})] that are independent of these quantities. It is not useless to recall that $\Delta f_{\pm}(k;t)$ is not the space Fourier transform of the chiral density $\Delta n_{\pm}(x,t)$, for it contains information also about the spatial correlations at different space points. Nevertheless, it can easily be checked that $\sum_k  \Delta f_{\pm}(k;t)=\int dx  \Delta n_{\pm}(x,t) =\Delta N_{\pm} \neq 0$, so that the total photoexcited charge in each branch is independent of the temperature and the chemical potential. The sum over all $k$'s yields the total photoexcited charge and is vanishing, $\sum_k  (\Delta f_{+}+\Delta f_{-})(k,t)=0$.

\subsection{Time correlations at a space point and the local tunneling density of states}
At a given space point $x$, electronic correlations  at different times are described by the matrix
\begin{eqnarray}
\lefteqn{\mathcal{G}(t_2,t_1;x)   \doteq  e^{-\frac{i{\rm e}}{\hbar}  \int_{t_2}^{t_1}   V(x^\prime,t)  \, d t^\prime} \times}  & &  \label{es-mat-def} \\& & \nonumber  \\
& \times &   
\left( \begin{array}{cc} \left\langle {\psi}_{+}^\dagger(x,t_1)   \psi_{+}(x,t_2)   \right\rangle \,  & \left\langle {\psi}_{+}^\dagger(x,t_1)   \psi_{-}(x,t_2)   \right\rangle \\
& \\
 \left\langle {\psi}_{-}^\dagger(x,t_1)   \psi_{+}(x,t_2) \right\rangle  & \left\langle  \psi_{-}^\dagger(x,t_1)  \psi_{-}(x,t_2)   \right\rangle \,  
\end{array} \right)   \, .  \nonumber
\end{eqnarray}
Similarly to the gauge-invariant density matrix Eq.(\ref{rho-mat-def}), the gauge invariance of the matrix (\ref{es-mat-def}) is ensured by the Wilson line (\ref{WL-def}). Note, however, that in this case of equal space points the phase pre-factor appearing in the first line (\ref{es-mat-def})  involves only the scalar potential $V$.\\ 

Again, for the present case of the Hamiltonian (\ref{H}), the~$\sigma_z$ projections are dynamically decoupled, and the matrix (\ref{es-mat-def}) is block-diagonal in the spinor basis,  with diagonal entries 
\begin{eqnarray}
\mathcal{G}_{\pm}(t_2,t_1;x) = \langle \psi^\dagger_{\pm}(x,t_1)\psi^{}_{\pm}(x,t_2)\rangle_\circ e^{-i\frac{{\rm e}}{\hbar}\int_{x_2}^{x_1} \!\!V(x, t^\prime) \, dt^\prime }     \nonumber \\
= \left\langle {\psi^\circ_{\pm}}^\dagger(x,t_1) {\psi^\circ_{\pm}}^{}(x,t_2) \right\rangle_\circ  \times \hspace{2cm} \nonumber \\
\times   e^{\mp i  (\phi_{\pm}(x,t_1) -\phi_{\pm}(x,t_2) \pm  \frac{{\rm e}}{\hbar}\int_{t_2}^{t_1} \!\!V(x, t^\prime) \, dt^\prime) } \,. \hspace{1cm}\label{nu_{pm}-def}
\end{eqnarray}
In particular, the effect induced by the electromagnetic field on the equilibrium correlation $\mathcal{G}_{\pm}^\circ$ is given by
\begin{eqnarray}
\lefteqn{  \Delta \mathcal{G}_{\pm}(t-\frac{t^\prime}{2},t+\frac{t^\prime}{2};x)  \doteq } \nonumber \\
&=&  \mathcal{G}_{\pm}(t-\frac{t^\prime}{2},t+\frac{t^\prime}{2};x) -\mathcal{G}^{\circ}_{\pm}(t-\frac{t^\prime}{2},t+\frac{t^\prime}{2};x)
=  \nonumber  \\
&=& \langle {\psi^\circ_{\pm}}^\dagger(x,t+\frac{t^\prime}{2}) {\psi^\circ_{\pm}}^{}(x,t-\frac{t^\prime}{2})  \rangle_\circ \! \left( e^{\mp i  \Delta \phi^{\rm es}_{\pm}(t,t^\prime;x) }    -1\right) \hspace{0.35cm} 
\label{Deltanu_{pm}-res-pre}
\end{eqnarray}
where we have introduced the average time $t=(t_1+t_2)/2$ and the time difference $t^\prime=t_1-t_2$, and the equal space gauge invariant phase difference
\begin{eqnarray}
\lefteqn{\Delta \phi^{\rm es}_{\pm}(t,t^\prime;x) \doteq } & & \label{Deltaphig-es-def} \\
& & \phi_{\pm}(x,t+\frac{t^\prime}{2}) -\phi_{\pm}(x,t-\frac{t^\prime}{2})  \pm  \frac{{\rm e}}{\hbar}\int_{t-\frac{t^\prime}{2}}^{t+\frac{t^\prime}{2}} \!\!\!V(x,  t^\prime) \, dt^\prime ,\nonumber
\end{eqnarray}
with $\phi_\pm$ given by Eqs.(\ref{phi+})-(\ref{phi-}). It can easily be checked with the use of Eq.(\ref{correlations0})  that, in the equal time limit $t^\prime  \rightarrow 0$, one recovers from Eq.(\ref{Deltanu_{pm}-res-pre}) the gauge invariant chiral densities [see fourth lines of (\ref{Deltan_{+}_res-gen})-(\ref{Deltan_{-}_res-gen})].\\

Time correlations are also suitably described in the frequency domain, by Fourier transforming Eq.(\ref{nu_{pm}-def}) with respect to times  
\begin{eqnarray}  
\lefteqn{ \mathcal{G}_\pm(\omega,\tilde{\omega};x)    = } & & \nonumber \\
& & =\frac{v_F}{L}\iint dt \,  dt^\prime e^{-i \omega t^\prime} e^{-i \tilde{\omega} t}\mathcal{G}_{\pm}(t-\frac{t^\prime}{2},t+\frac{t^\prime}{2};x) \,\,. \label{nu_{pm}-def-omega-space}
\end{eqnarray}
By substituting Eq.(\ref{nu_{pm}-def}) into Eq.(\ref{nu_{pm}-def-omega-space}) and by proceeding in a similar way as for the density matrix, one obtains
\begin{eqnarray}
\lefteqn{\mathcal{G}_{\pm}(\omega,\tilde{\omega};x)  = \frac{\pi\delta(\tilde{\omega})}{L} \, \mp \frac{v_F}{2 l_T L}  \int e^{-i \tilde{\omega} t} dt } & &   \label{nu_{pm}-fin}  \\
& & \hspace{1cm} \times \int_{-\infty}^{+\infty}\frac{\sin\left(\Delta\phi^{\rm es}_{\pm}(t^\prime,t;x)   \pm  (\omega -\mu/\hbar) t^\prime  \right)}{\sinh[\pi v_F t^\prime/l_T]}\, dt^\prime \,  \nonumber
\end{eqnarray}
whose limit $\tilde{\omega} \rightarrow 0$ yields the local tunneling density of states, i.e. $\nu(\omega;x) \doteq \mathcal{G}_{\pm}(\omega,\tilde{\omega}=0;x)$.\\
In particular, the photoexcited local tunneling density of states (LDOS) is
\begin{eqnarray}
\lefteqn{\Delta \nu_{\pm}(\omega;x) \doteq \nu_{\pm}(\omega;x) -\nu^\circ_{\pm}(\omega;x) = } & & \nonumber \\
&=&  \mp \frac{v_F}{2 L l_T}    \int_{-\infty}^{+\infty} \!\!\! dt\,  \int_{-\infty}^{+\infty}  \, dt^\prime \,  \label{fk_{pm}-fin} \\
& & \times \,\frac{\sin\left[\Delta\phi^{\rm es}_{\pm}(t^\prime,t;x)   \pm  (\omega -\mu/\hbar) t^\prime  \right]\mp \sin\left[(\omega -\mu/\hbar) t^\prime \right]}{\sinh[\pi v_F t^\prime/l_T]}\,   .  \nonumber 
\end{eqnarray}
Just like the momentum distribution $\Delta f(k;t)$ in Eq.(\ref{Deltafk_{pm}-fin}), the photoexcited  LDOS $\Delta \nu(\omega;x)$ depends on the temperature $T$ and on the chemical potential $\mu$, in striking contrast with the photoexcited density profiles $\Delta n_{\pm}(x,t)$ [see Eqs.(\ref{Deltan_{+}_res-gen})-(\ref{Deltan_{-}_res-gen})] that are independent of these quantities.
\section{The case of a localised electric pulse}
\label{sec-4}
We shall now apply the general results obtained in the previous section to the case of an electric pulse that is applied for a finite duration $\tau$ and is localised over a region of size~$\Delta$, with $\Delta$ not necessarily equal to the longitudinal length $L$ of the QSH edge channels.  
We start by considering a monochromatic radiation with frequency $\Omega$, with a sharp cut-off in space and time, i.e. 
\begin{equation}\label{Eplanewave-cutted}
E(x,t)=E_0   \cos(\frac{\Omega x}{c}) \cos(\Omega t) \, \theta(\frac{\Delta}{2}-|x|)   \,\theta(\frac{\tau}{2}-|t|)  
\end{equation}
where $\theta$ is the Heaviside function.
Although not very realistic, the form (\ref{Eplanewave-cutted}) allows one to  qualitatively illustrate the effects of the spatial and temporal confinements $\Delta$ and $\tau$, since   simple expressions for the photoexcited density profiles are straightforwardly obtained upon substituting Eq.(\ref{Eplanewave-cutted}) into the first two lines of Eqs.(\ref{Deltan_{+}_res-gen})-(\ref{Deltan_{-}_res-gen}). 

The conventional far field regime is obtained in the limits of a long ($\Omega \tau \gg 1$) and spatially extended pulse $\Delta/L \gg 1$, where the whole electron system is exposed to the radiation  for many oscillation periods, so that both energy and momentum conservation constraints hold for the excited electrons and the absorbed `photon'. Because the velocities $v_F$ and $c$ of the electronic and photonic spectra are different, these  constraints cannot be both fulfilled, and a vanishing intra-branch  response is obtained,
\begin{eqnarray}
\Delta n_{\pm}(x,t) \propto \displaystyle \pm \frac{{\rm e}E_0 c}{2\hbar} \,   \delta(\Omega v_F- \Omega c)    =0 
 \quad. \label{Deltan_{pm}-planewave-far-field}
\end{eqnarray} 
However, when the time or/and the space confinement is finite, either of these constraints or both are relaxed and a non-vanishing intra-branch optical transition is possible. In particular, if the radiation is applied   everywhere ($\Delta \gg L$) but for a short time ($\Omega \tau \ll 1$), the energy conservation constraint is relaxed and one obtains  
\begin{equation}
\Delta n_{\pm}(x,t) = \displaystyle \pm \frac{{\rm e}E_0 \tau}{2\pi \hbar} \,   \cos \left[  \frac{\Omega}{c} (x \mp v_F  t) \right]   \quad. \label{Deltan_{pm}-planewave}
\end{equation}
In this regime only the transferred momentum $q=\Omega/c$ is conserved, while the electron density oscillates in time with a frequency $\Omega_{el}=\Omega v_F/c$ lower than the electromagnetic wave. Typically one has $\Omega L/c \ll 1$, where $L$ is the length for the QSH edge system, i.e. the electron system `sees' the electromagnetic wave as a uniform electric field oscillating in time. Note that the amplitude of the photoexcited electron densities is governed by the product $E_0 \tau$.

In the opposite case of a radiation applied for a long time ($\tau \rightarrow \infty$) but over a short spatial region~$\Omega \Delta/v_F \ll 1$ one obtains, away from the exposed region ($|x|> \Delta/2$),
\begin{equation}
\Delta n_{\pm}(x,t) = \displaystyle \pm \frac{{\rm e}E_0 \Delta}{2\pi \hbar v_F} \,   \cos \left[  \frac{\Omega}{v_F} (x \mp v_F  t) \right]   \quad. \label{Deltan_{pm}-planewave}
\end{equation}
In this case the  energy is conserved, so that the electron density  oscillates in time with the {\it same} frequency $\Omega$ as the electromagnetic wave, while the momentum conservation constraint is relaxed, and the electron wave vector $q_{el}=\Omega/v_F > q_{light}=\Omega/c$ is higher than the radiation field.
Notice that, even if typically $\Omega L/c \ll 1$, $\Omega L/v_F$ is not necessarily small. In this regime the electron system `sees' the electromagnetic wave as a localised time-dependent gate voltage   that oscillates with the frequency $\Omega$, and the amplitude of the photoexcited electron densities is governed by the product $E_0 \Delta/v_F$.\\

In the case of a localised electric pulse, the spatial and temporal confinement interplay, as we shall now describe with the more realistic case of a gaussian pulse 
\begin{equation}\label{Egauss-cos}
E(x,t)=E_0 \, e^{-\frac{x^2}{2 \Delta^2}} \, e^{-\frac{t^2}{2 \tau^2}} \cos(\Omega t)\quad,
\end{equation}
where $\Delta$ and $\tau$ denote the standard deviation around the space and time origin, respectively. Here below we present the result for the case (\ref{Egauss-cos}), focussing on the density space profile  and the momentum distribution of the photoexcited wave packets.\\

\subsection{Photoexcited density profiles}
Substituting the pulse (\ref{Egauss-cos}) into Eqs.(\ref{Deltan_{+}_res-gen})-(\ref{Deltan_{-}_res-gen}), one obtains
\begin{eqnarray}
\lefteqn{\displaystyle  \Delta n_{\pm}(x,t)=  \pm \frac{{\rm e} E_0}{2 \pi \hbar}\frac{D}{v_F} \sqrt{\frac{\pi}{2}}e^{-\frac{\Omega^2 D^2}{2 v_F^2}} e^{-\frac{(x \mp v_F t)^2}{2(\Delta^2+(v_F \tau)^2)} } \times }  \label{Deltan_{pm}-Egauss-pulse} & & \\
& &  \Re \left\{ e^{i \frac{\Omega D^2}{v_F \Delta^2}(x \mp v_F t)} \left[ 1+{\rm Erf}\left(\frac{D}{\sqrt{2}}\left(\pm \frac{x}{\Delta^2}+\frac{t}{v_F \tau^2}\pm \frac{i \Omega}{v_F}\right)\!\right)\right]\right\}     
\nonumber
\end{eqnarray}
where
\begin{equation}\label{D-def}
D  \doteq  \left( \frac{1}{\Delta^2}+\frac{1}{(v_F \tau)^2}\right)^{-1/2} = \frac{v_F \tau \Delta}{\sqrt{\Delta^2+(v_F \tau)^2}}
\end{equation}
is an effective length scale involving both the space extension $\Delta$ and the time duration $\tau$ of the pulse. Let us now analyze some specific limits of the expression (\ref{Deltan_{pm}-Egauss-pulse}).
\subsubsection{Low frequency limit} 
Since the spectrum of the QSH is gapless, optical transitions are possible also in the limit of low frequency $\Omega$. Indeed, for $\Omega \tau \ll 1$ and $\Omega \Delta/v_F \ll 1$, Eq.(\ref{Deltan_{pm}-Egauss-pulse}) reduces to
\begin{eqnarray}
\lefteqn{\Delta n_{\pm}(x,t)=\pm \displaystyle  \frac{{\rm e} E_0}{2 \pi \hbar}\frac{D}{v_F} \sqrt{\frac{\pi}{2}} e^{-\frac{(x \mp v_F t)^2}{2(\Delta^2+T^2)} }   } \hspace{1cm}\label{Deltan_{pm}-Egauss-pulse-low} \\ & &  \times \left\{ 1+{\rm Erf}\left[\frac{D}{\sqrt{2}}\left(\pm\frac{x}{\Delta^2}+\frac{v_F t}{T^2} \right)\right]\right\} \nonumber   
\end{eqnarray}
\noindent providing useful physical insights. During the finite duration of the pulse (i.e. for $|t| \lesssim \tau$), the densities $\Delta n_{\pm}(x,t)$ do not evolve, in general, as right- and left-movers. In contrast, after the pulse, i.e. for times $t \gg \tau$, Eq.(\ref{Deltan_{pm}-Egauss-pulse-low}) reduces to 
\begin{equation}
\Delta n_{\pm}(x,t)  \simeq   \displaystyle  \pm \frac{{\rm e} E_0}{2 \pi \hbar}\frac{\tau \Delta\sqrt{2 \pi}}{\sqrt{\Delta^2+(v_F \tau)^2}}  e^{-\frac{(x \mp v_F t)^2}{2(\Delta^2+(v_F \tau)^2)} } \, , \label{Deltan_{pm}-Egauss-pulse-zero-omega-long-times}
\end{equation}
which describes two photoexcited wave packets propagating rightwards and leftwards, respectively. Notably, although the shape of the electron densities $\Delta n_{\pm}$ is gaussian like the applied pulse (\ref{Egauss-cos}), their space extension  
\begin{equation}
\Delta_{el}=\sqrt{\Delta^2+(v_F \tau)^2}
\label{Delta-el}
\end{equation}
is determined by both the space extension $\Delta$ and the time duration $\tau$ of the electric pulse, and is bigger than $\Delta$. In the particular limit of a short pulse, $\tau \ll \Delta/v_F$, Eq.(\ref{Egauss-cos}) can be treated as a $\delta$-pulse, $E(x,t)= \mathcal{E}_0 \, \delta(t)\exp(- x^2/2 \Delta^2)$   upon identifying $\mathcal{E}_0=E_0 \tau \sqrt{2\pi}$, and the profile of the photoexcited electron density has the same space extension $\Delta$ as the pulse.\\

\begin{figure}[h]
\centering
\includegraphics[width=\columnwidth]{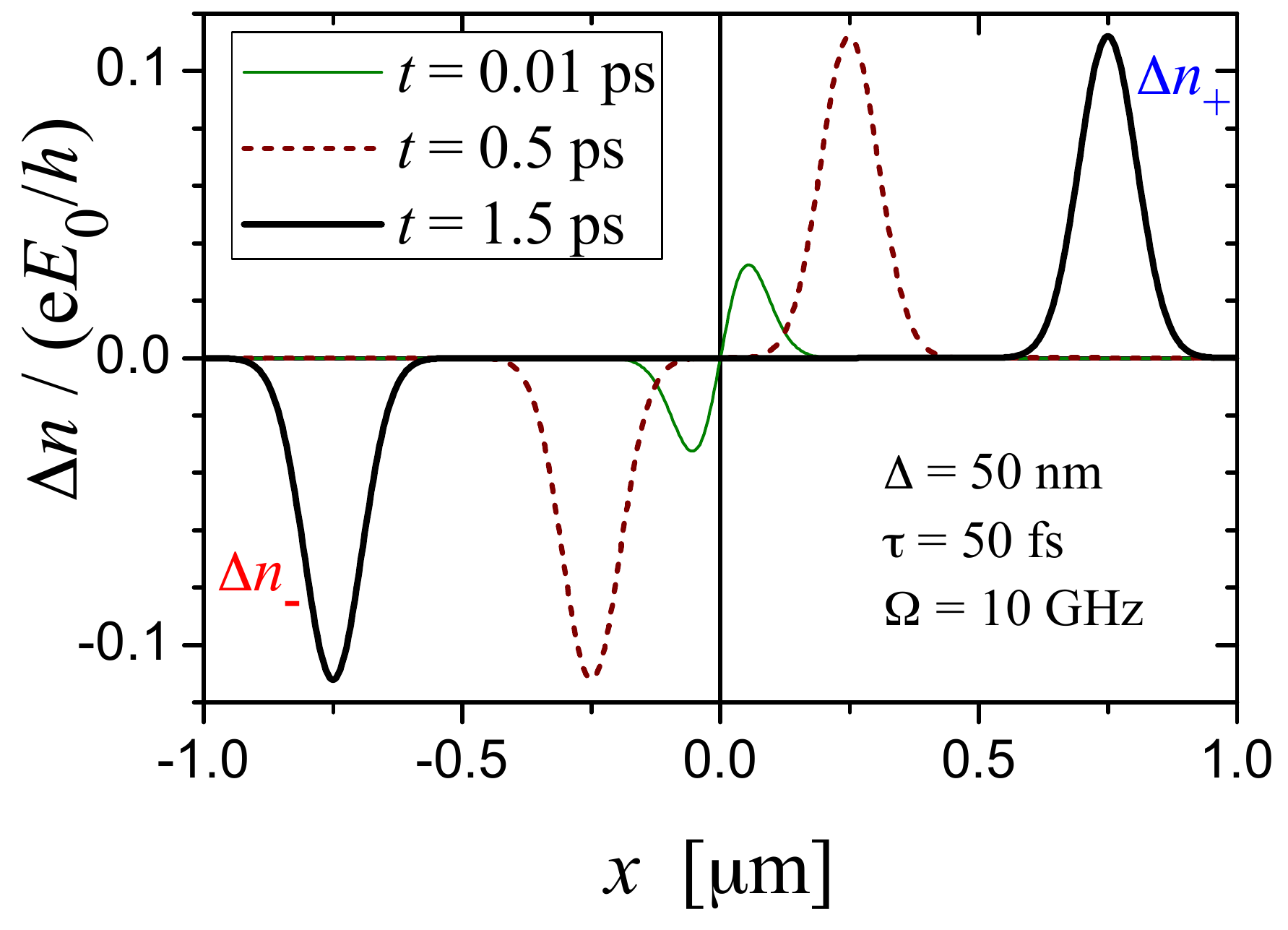}
\caption{
(Color online) The space profile of the electron density $\Delta n=\Delta n_{+}+\Delta n_{-}$ photoexcited by the pulse (\ref{Egauss-cos}) in the regime of low frequency, namely for $\Omega=10 {\rm GHz}$, $\tau=50 \, {\rm fs}$  and $\Delta=50 \, {\rm nm}$, is shown at three different snapshots: $t=0.01 \,{\rm ps}$ (thin solid curve), $t=0.5 \,{\rm ps}$ (dashed curve) and $t=1.5 \, {\rm ps}$ (thick solid curve).  In this regime, the electronic density profiles reproduces the space profile of the electric pulse, with, however, a renormalised space extension parameter $\Delta_{el}$ given by Eq.(\ref{Delta-el}).  The profile is independent of the temperature and chemical potential of the initial electronic equilibrium state, and is linear in the amplitude $E_0$ of the applied pulse, as is emphasized by the vertical axis label. The two photoexcited wave packets counter propagate with velocity $v_F=5 \times 10^5 {\rm m/s}$.
}
\label{Fig-dens-low-frequency}
\end{figure}

Figure \ref{Fig-dens-low-frequency} shows the process of electron wave packets photoexcitation and propagation for the low frequency case, specifically for $\Omega=10 {\rm GHz}$, $\tau=50 {\rm fs}$ and  $\Delta=50 \, {\rm nm}$ in Eq.(\ref{Egauss-cos}), through a series of snapshots of the total density $\Delta n=\Delta n_{+}+\Delta n_{-}$. As one can see, during the application of the electric pulse (thin solid curve) the density is gradually modified and starts to display its two components $\Delta n_{+}$ and $\Delta n_{-}$  with opposite sign, which counter propagate with velocity $v_F$ without dispersion once the pulse has ended (dashed and thick solid curves). In this case the wave packet profile is gaussian, with a spatial extension given by Eq.(\ref{Delta-el}). 
\subsubsection{Finite frequency: asymptotic behavior} 
\begin{figure}[h]
\centering
\includegraphics[width=\columnwidth]{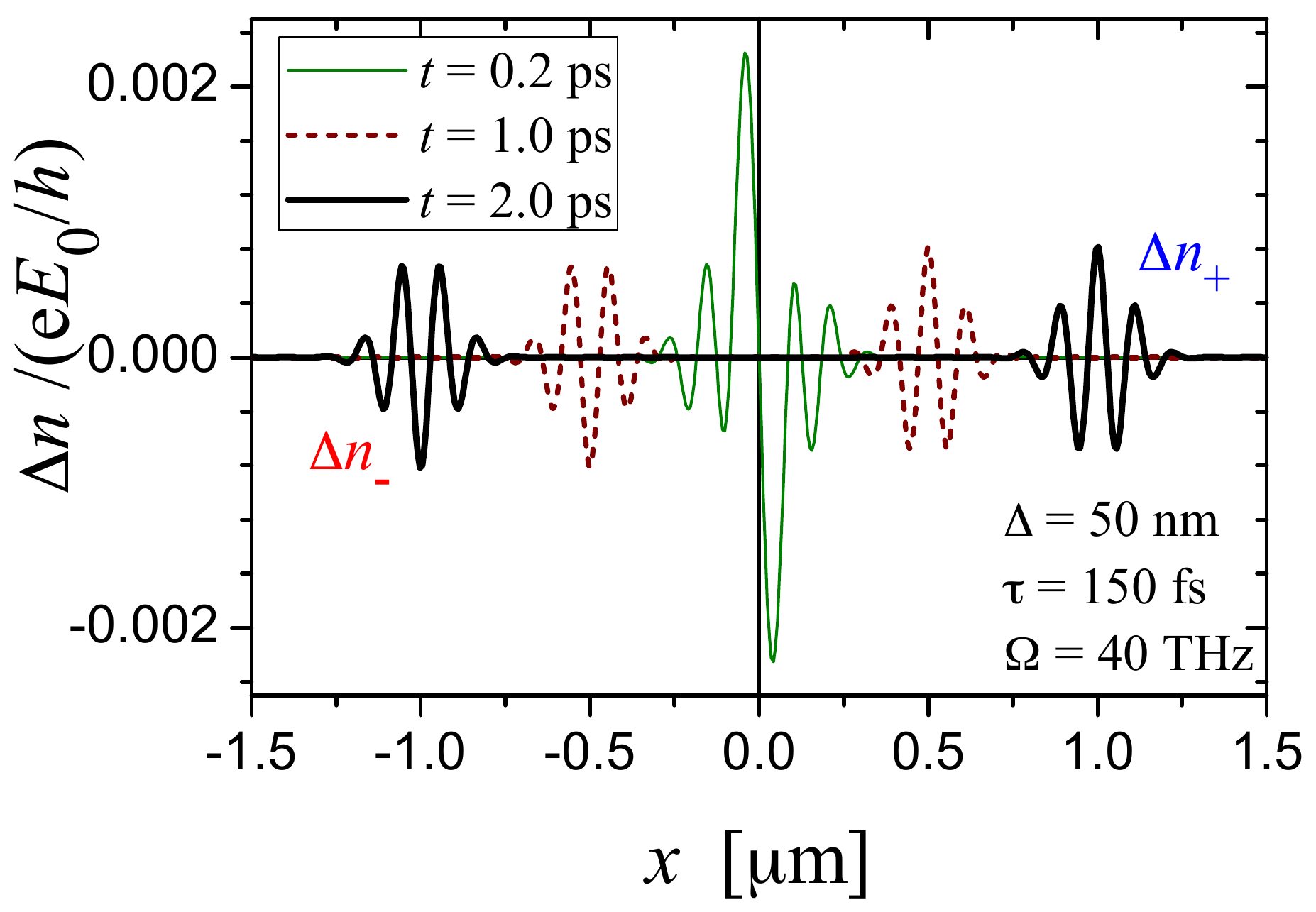}
\caption{
(Color online) The space profile of the electron density $\Delta n=\Delta n_{+}+\Delta n_{-}$ photoexcited by the pulse (\ref{Egauss-cos}) in the regime of high frequency, namely for $\Omega=40 {\rm THz}$, $\tau=150 \,{\rm fs}$  and $\Delta=50 \, {\rm nm}$, is shown at three different snapshots: $t=0.2 \, {\rm ps}$ (thin solid curve), $t=1.0 \, {\rm ps}$ (dashed curve) and $t=2.0 \, {\rm ps}$ (thick solid curve). In this regime, the electronic wave packet exhibits space and time oscillations, characterised by a  frequency $\Omega_{el}$, given by Eq.(\ref{omega-el}), renormalised with respect to the pulse frequency $\Omega$.
Again, the profile is independent of the temperature and chemical potential of the initial electronic equilibrium state, and is linear in the amplitude $E_0$ of the applied pulse. The propagation velocity is  $v_F=5 \times 10^5 {\rm m/s}$. 
}
\label{Fig-dens-hih-frequency}
\end{figure}

Let us now consider a finite frequency $\Omega$, and analyze the asymptotic behavior for long times and/or positions. More specifically, for  
$
\left|D \left(\frac{x}{\Delta^2}+\frac{t}{v_F \tau^2}\pm \frac{i\Omega}{v_F}\right)  \right|  \, \, \gg 1
$, Eq.(\ref{Deltan_{pm}-Egauss-pulse}) reduces to
\begin{eqnarray}
\displaystyle \Delta n_{\pm}(x,t)  &= &  \pm \frac{{\rm e} E_0}{2 \pi \hbar} \,  \frac{\tau \Delta\sqrt{2 \pi}}{\sqrt{\Delta^2+(v_F \tau)^2}}     \,   \, e^{-\frac{\Omega^2 D^2}{2v_F^2}} e^{-\frac{(x \mp v_F t)^2}{2 (\Delta^2+(v_F \tau)^2)}}   \nonumber \\
& & \times \cos{\left[ \frac{\Omega \, v_F \tau^2}{\Delta^2 +(v_F \tau)^2} (x \mp v_F t)\right]}  \, \, .\label{phi_{pm}-Egauss-long} 
\end{eqnarray}
The finite frequency yields an exponential suppression $\exp(-\Omega^2 D^2/2v_F^2)$ of the electric pulse amplitude $E_0$, where $D$ is given by Eq.(\ref{D-def}).
Furthermore, in this case the photoexcited electron densities exhibit, besides the gaussian envelope, an oscillatory behavior characterised by a frequency 
\begin{equation}\label{omega-el}
\Omega_{el}=\Omega \frac{(v_F \tau)^2}{\Delta^2 +(v_F \tau)^2}
\end{equation}
that depends on both the finite space extension $\Delta$ and time duration $\tau$ of the electric pulse, and that is lower than the frequency $\Omega$ of the applied pulse. These features are described in Fig.\ref{Fig-dens-hih-frequency}, which shows the photo excitation of wave packets in the high frequency regime, specifically for $\Omega=40 {\rm THz}$, $\tau=150{\rm fs}$, and $\Delta=50{\rm nm}$.\\  These results show how the photoexcited charge space profile can be tailored by the applied pulse parameters.

\subsection{Photoexcited momentum distribution}
The photoexcited momentum distribution $\Delta f(k;t)$ can be obtained from Eq.(\ref{Deltafk_{pm}-fin}) and depends on the equal time gauge invariant phase difference $\Delta \phi^{\rm et}_{\pm}(x,y;t)$ in Eq.(\ref{Deltaphig-et-def}). 
The latter can be evaluated in any gauge $(V,A)$ yielding the gaussian electric pulse (\ref{Egauss-cos}); two examples are noteworthy among all possible gauges, namely a gauge with purely scalar potential,
\begin{equation}\label{V-gauss-pulse}
\begin{array}{lcl}
V(x,t)&=& \displaystyle -E_0\Delta \sqrt{\frac{\pi}{2}} \, {\rm Erf}\left(\frac{x}{\sqrt{2}\Delta}\right)   e^{-\frac{t^2}{2 \tau^2}}    \cos(\Omega t)  \\
A(x,t) &=& 0
\end{array} 
\end{equation}
and a gauge with purely vector potential,
\begin{equation}\label{A-gauss-pulse}
\begin{array}{lcl}
V(x,t)&=& 0 \\
A(x,t) &=& \displaystyle -c \, E_0 \,\tau \sqrt{\frac{\pi}{2}}\, e^{-\frac{x^2}{2 \Delta^2}} \, e^{-\frac{(\Omega \tau)^2}{2}}  \\
& & \times \left\{1+\Re  \left[{\rm Erf}\left(\frac{1}{\sqrt{2}}(\frac{t}{\tau}+i \Omega \tau) \right) \right]\,   \right\}   \quad,
\end{array}
\end{equation}
where ${\rm Erf}$ denotes the error function.  In both cases, using Eqs.(\ref{phi+}), (\ref{phi-}) and (\ref{Deltaphig-et-def}), one finds
for the gaussian pulse~(\ref{Egauss-cos}) 
\begin{eqnarray}
\lefteqn{\Delta \phi^{\rm et}_{\pm}(x,y;t)        
= \pm  \frac{{\rm e} E_0}{\hbar} \Delta \sqrt{\frac{\pi}{2}}  \int_{-\infty}^t  dt^\prime \, e^{-\frac{{t^\prime}^2}{2 \tau^2}}  \, \cos(\Omega t^\prime)\, \times\,  \, } \, \,& & \label{phipm-et-Egauss-pulse}   \\
& &   \left\{ {\rm Erf}\left[\frac{x+\frac{y}{2}\mp v_F (t-t^\prime)}{\sqrt{2}\Delta}\right]  -{\rm Erf}\left[\frac{x-\frac{y}{2}\mp v_F (t-t^\prime)}{\sqrt{2}\Delta}\right]    \right\} , 
\nonumber
\end{eqnarray}
and a numerical integration of Eq.(\ref{Deltafk_{pm}-fin}) straightforwardly enables one to determine the dependence of $\Delta f_\pm(k;t)$ on the amplitude $E_0$ of the electric pulse and the temperature $T$. In particular, for times longer than the pulse duration, $t \gg \tau$, the momentum distributions $\Delta f_{\pm}(k;t)$ of the two counter propagating wave packets shown in Figs.\ref{Fig-dens-low-frequency} and \ref{Fig-dens-hih-frequency} turn out to be asymptotically independent of time. Exploiting the spatial  symmetry of  the electric pulse (\ref{Egauss-cos}), the simple relation $\Delta f_{-}(k;t)|_{k_F}=-\Delta f_{+}(k;t)|_{-k_F}$ can be obtained. We shall thus focus on the case right-moving electron wave packet.  

\begin{figure}[h]
\centering
\includegraphics[width=\columnwidth]{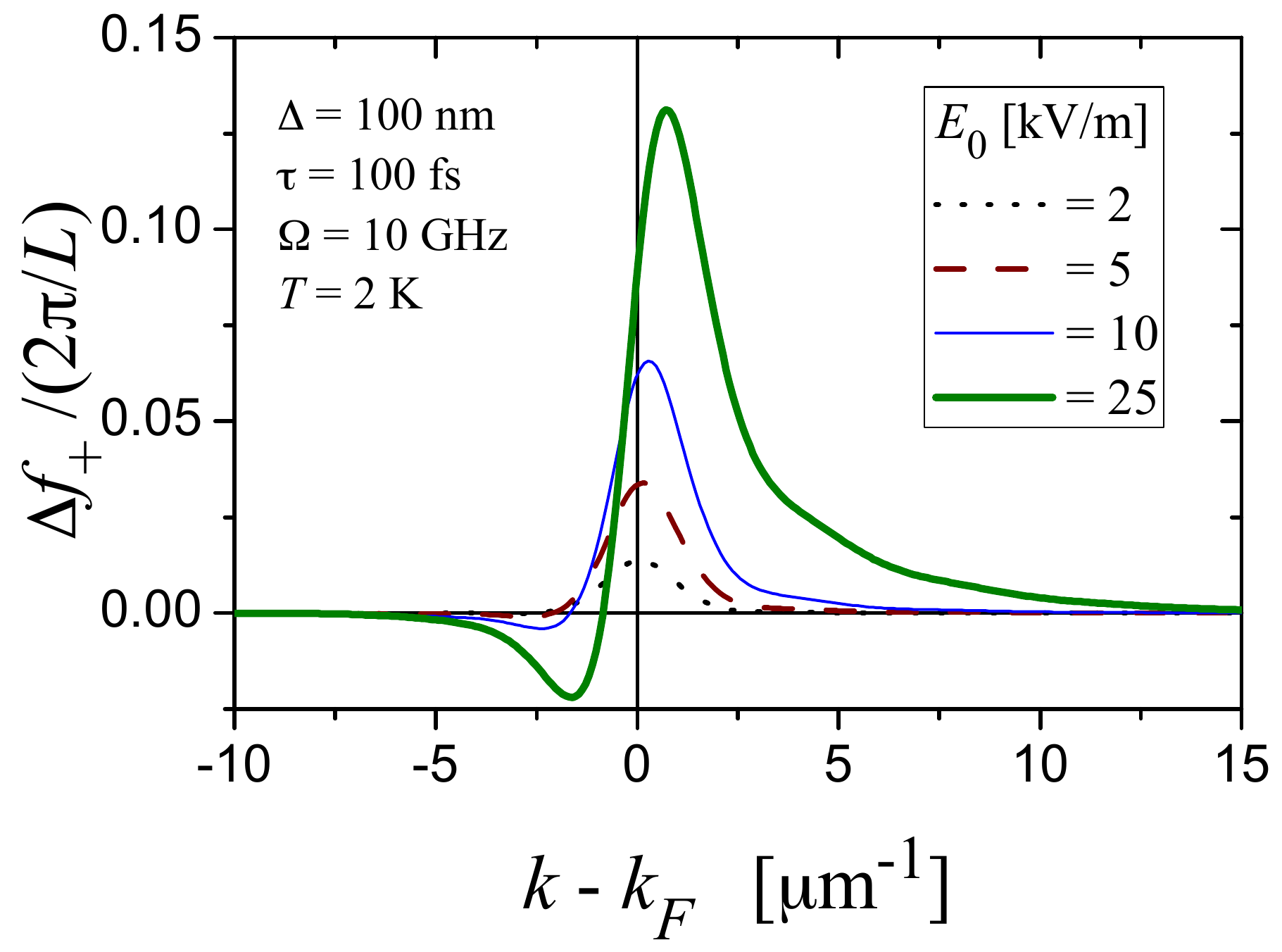}
\caption{
(Color online) The momentum distribution $\Delta f_{+}(k)$ of the right-moving electron wave packet photoexcited by the pulse (\ref{Egauss-cos})  with $\Delta=100 \, {\rm nm}$ and $\tau = 100\, {\rm fs}$ and $\Omega=10 {\rm GHz}$  is shown as a function of the wave vector deviation $k-k_F$ from the Fermi wave vector $k_F$, for different values of the electric pulse amplitude $E_0=2 {\rm kV/m}$ (dotted curve), $E_0=5 {\rm kV/m}$ (dashed curve), $E_0=10 {\rm kV/m}$ (thin solid curve), and   $E_0=25 {\rm kV/m}$ (think solid curve).  The initial equilibrium state is characterised by a temperature $T=2 {\rm K}$ and a chemical potential $\mu= \hbar v_F k_F$, with $v_F=5 \times 10^5 {\rm m/s}$. Differently from the space profile of the photoexcited density shown in Figs.~\ref{Fig-dens-low-frequency} and~\ref{Fig-dens-hih-frequency}, $\Delta f_{+}(k)$ depends non linearly on $E_0$. 
}
\label{Fig-fk-EoVAR}
\end{figure}

Figure~\ref{Fig-fk-EoVAR} shows the photoexcited  momentum distribution $\Delta f_{+}(k;t)$ in the low frequency regime,  for a fixed value of the temperature and chemical potential of the initial equilibrium state, and for various values of the amplitude $E_0$ of the applied pulse. Differently from the spatial density profile $\Delta n_\pm(x,t)$ shown above, $\Delta f_{+}(k;t)$ does not rescale simply linearly with $E_0$. In particular, while for weak fields $\Delta f_{+}(k;t)$ is roughly symmetric around the Fermi wavevector $k_F$, for stronger fields it exhibits a negative dip  below the Fermi level and a broader positive peak above it; notably, the integral over $k$ is non vanishing, and increases with $E_0$. As observed above, for massless Dirac electrons the electric pulse effectively `creates' electrons in the chiral branch e.g. $+$, while `depleting' the other branch $-$, the total created charge remaining of course vanishing. 

\begin{figure}[h]
\centering
\includegraphics[width=\columnwidth]{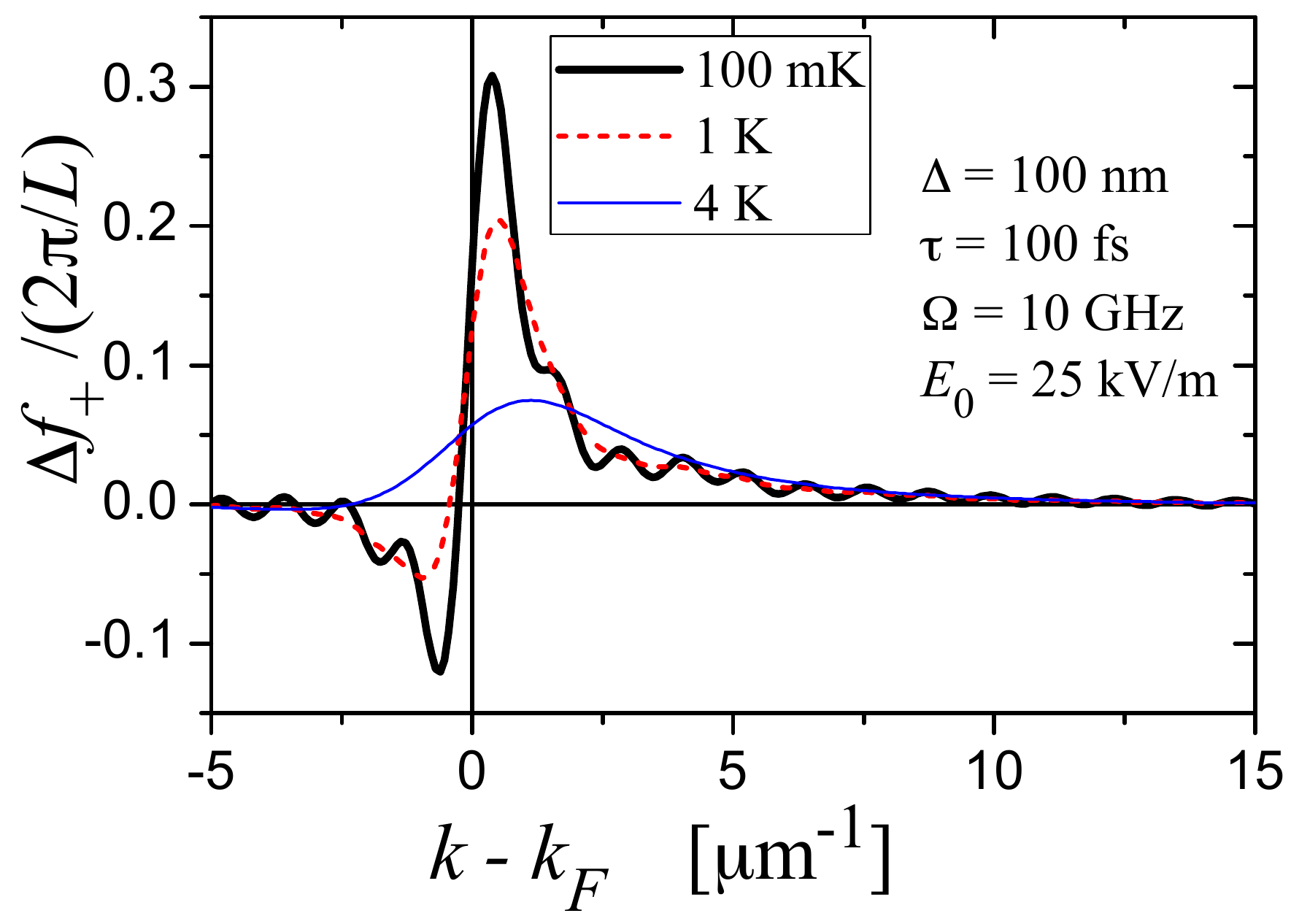}
\caption{
(Color online) The momentum distribution $\Delta f_{+}(k)$ of the right-moving electron wave packet photoexcited by the pulse (\ref{Egauss-cos})  with $E_0=25 {\rm kV/m}$, $\Delta=100 \, {\rm nm}$ and $\tau = 100\, {\rm fs}$ and $\Omega=10 {\rm GHz}$  is shown as a function of the wave vector deviation $k-k_F$ from the Fermi wave vector $k_F$, for different values of temperature $T$ of the  initial equilibrium state: $T=100 \,{\rm mK}$ (thick solid curve), $T=1 \,{\rm K}$ (dashed curve) and $T=4 \,{\rm K}$ (thin solid curve). The Fermi velocity is $v_F=5 \times 10^5 {\rm m/s}$. While sharp peaks and oscillations related to the localisation length scale $\Delta$ of the applied pulse arise at low temperature, with increasing $T$ the oscillations are washed out by thermal fluctuations and the peaks of $\Delta f_{+}(k)$ decrease and  broaden. 
}
\label{Fig-fk-TVAR}
\end{figure}

Figure~\ref{Fig-fk-TVAR} shows the strong temperature dependence of  $\Delta f_{+}(k)$. In particular, at low temperature $\Delta f_{+}(k)$ displays sharp dips and peaks with an oscillatory pattern, as a result of the spatial localisation $\Delta$ of the applied electric pulse. Indeed for a uniform field $E_0$ applied for a duration~$\tau$, the photoexcited distribution would simply be  $\Delta f_{+}(k)=f^\circ_{+}(k-{\rm e}E_0 \tau/\hbar)-f^\circ_{+}(k)$, without oscillations, as can be obtained from the general result Eq.(\ref{Deltafk_{pm}-fin}). For higher temperature values the oscillations of $\Delta f_{+}(k)$ are washed out by thermal fluctuations and the dips and peaks  decrease and  broaden. It is also worth mentioning that, as a further effect of the pulse spatial localisation, off-diagonal entries ($k_1 \neq k_2$) arise in the momentum representation of the density matrix $\rho(k_2,k_1,t)$, which is instead purely diagonal for a uniform electric pulse.  

These results explicitly show the striking difference of the  photoexcitation effects  in $k$-space as compared to real space: with varying the initial temperature $T$ and the amplitude $E_0$ of the applied pulse, the occupation of electronic states in $k$-space is modified in a complex and non-linear way, which, however, leaves the spatial electron profile $\Delta n(x,t)$ unaltered by $T$ and linearly rescaled by $E_0$ (see Figs.~\ref{Fig-dens-low-frequency} and \ref{Fig-dens-hih-frequency}). 
As observed at the end of Sec.\ref{sec-3B}, such simple behavior in real space is a signature of the chiral anomaly. In turn,  the relations $\int \Delta n_{\pm}(x,t) dx= \sum_k \Delta f_\pm(k,t)$ imply that, despite that $\Delta f_\pm(k,t)$ depends on $T$, its integral over $k$ does not. 

\section{Discussion and Conclusion}
\label{sec-5}
We have investigated the photoexcitation of electron wave packets in QSH helical edge states, described by a massless Dirac fermion Hamiltonian, exposed to an electric pulse applied along the QSH edge. 
In massless Dirac fermions the response to an electromagnetic field involves a non-trivial phenomenon: neither does the field directly couple the two helical branches, nor it simply redistribute electronic states within each branch. Instead, it effectively `creates' electrons in one branch and `depletes' electrons in the other branch accordingly, leading to an inter-branch transfer of electrons occurring via the inner depths of the Dirac Sea. This subtle effect, known as the chiral anomaly, breaks the conservation laws (\ref{cont-right}) and (\ref{cont-left}) that would be expected to hold for each chiral branch, on the basis of the Hamiltonian symmetries.   
We have fully taken into account these aspects by deriving the exact quantum dynamics of the electron field operator, and by computing electron densities and correlations with a regularisation procedure that ensures the gauge invariance of the results via the inclusion of a suitable Wilson line. 

Our results show that, while for an applied radiation in the far field regime electric dipole transitions are forbidden by helical selection rule and only transitions involving magnetic Zeeman coupling or bulk states are allowed, when the electric pulse is localised over a finite length and has a finite duration, the photoexcitation of electron wave packets is possible as a result of purely electrical intra-branch transitions in the edge states. In particular, we have shown that during the application of the pulse the helical components lose their character of right- and left-movers, despite their mutually decoupled dynamics. In contrast, after the ending of the pulse, the photoexcited wave packets propagate in opposite directions maintaining both their spin orientation and their spatial profile without dispersion (see Figs.~\ref{Fig-dens-low-frequency} and \ref{Fig-dens-hih-frequency}), as a result of the helical nature and the linearity of the Dirac spectrum. 
We have computed both the electron  space correlations at equal times, and the time correlation at a fixed space point. In particular, for the case of a gaussian electric pulse (\ref{Egauss-cos}), we have discussed in detail how the momentum distributions $\Delta f_\pm(k;t)$ of the photo excited wave packets depend on the temperature $T$ and the chemical potential $\mu$ of the initial electronic equilibrium state (see Fig.~\ref{Fig-fk-EoVAR}), and we have shown its  non-linear behavior as a function of the amplitude $E_0$ of the applied field (see Fig.~\ref{Fig-fk-TVAR}).

Importantly, we have proven that the space profile $\Delta n_{\pm}(x,t)$ of the photoexcited wave packets is instead {\it independent} of $T$ and $\mu$, and is determined {\it uniquely} and {\it linearly} by the applied electric pulse $E(x,t)$  [see Eqs.(\ref{Deltan_{+}_res-gen})-(\ref{Deltan_{-}_res-gen})]. This is a  signature of the chiral anomaly in 1D massless Dirac fermions. 
Indeed the term on the right hand side of Eq.(\ref{cont-right-left-new}), which breaks the chiral conservation laws and is responsible for a non-vanishing response to the applied electric field, depends only on the universal constant ${\rm e}/2\pi \hbar$ and the electric field $E(x,t)$ itself, in a linear manner,  and not on the electronic degrees of freedom. 

The search for signatures of the chiral anomaly is currently on the spotlight in condensed matter physics, due to the discovery of 3D Weyl semimetals\cite{chen2014,hasan2015a,hasan2015b,ong2015}, where the anomalous term depends on both the electric and magnetic field and is expected to lead to an unconventional electron pump. 
Quite recently its impact has also been envisaged in 1D QSH edge states, and a system of two QSH quantum dots has been proposed to observe its signatures in real space\cite{trauzettel2016}. In this respect, not only our results indicate a signature of the chiral anomaly in QSH edge states, they also have a practical consequence, since the shape of the propagating wave packets is shown to be tailored through the applied electric pulse only. Indeed its spatial extension $\Delta_{el}$ and frequency $\Omega_{el}$ after the pulse ending have been shown to depend, at low as well as at finite frequencies,  on a combination of both the space extension $\Delta$ and the time duration $\tau$ of the pulse [see Eq.(\ref{Delta-el})  and Eq.(\ref{omega-el})].\\
The results presented here are valid in the mesoscopic regime. Before concluding, however, we would like to discuss the possible impact of a few aspects that we have neglected in our analysis, and a possible scheme for implementation of the proposed setup. 

{\it Electron-phonon coupling effects.} 
As observed above, after the ending of the pulse, the photoexcited density profiles $\Delta n_\pm(x,t)$ are shown to propagate without dispersion, as a consequence of the linear spectrum of the massless Dirac fermions. While elastic scattering off non-magnetic impurities is forbidden by topological protection, inelastic electron-phonon coupling may in principle affect this ideal scenario. For the case of metallic SWNTs --which are also characterised by a linear spectrum-- a recent study has shown that modifications to the dispersionless propagation mainly arise from electron-phonon backscattering terms, due  in that case  to breathing phonon modes that couple the two graphene sublattices\cite{rosati2015}. However, such backscattering terms have no counter part in the QSH edge states, as electron-phonon coupling does not mix spin species. Yet, electron-phonon coupling may interplay with Rashba impurities, which do allow spin-flip processes when combined with a momentum reversal, leading to inelastic two-particle backscattering that would affect the wave packet propagation. It should be mentioned, however,  that the backscattering current, although in principle present, has been evaluated as extremely negligible at low applied voltage or temperature, maintaining the QSH edge state topological protection  de facto. \cite{budich2012} 

{\it  Electron-electron interaction effects.}  
The present analysis has neglected electron-electron interaction in the QSH edge states. As is well known,  interactions in 1D electron systems typically induce a Luttinger liquid behavior, leading the  correlation functions to exhibit a non-analytic behavior characterised by a non-universal Luttinger parameter $K \le 1$, which depends on the interaction screened by the substrate. The effective interaction strength has been predicted to be weak in HgTe/CdTe quantum wells\cite{teo-kane2009}, and stronger in InAs/GaSb quantum well\cite{zhangPRL2009}. However, it should be pointed out that, at the moment, experimental evidence for a Luttinger liquid behavior in QSH edge states is extremely limited\cite{du2015}. Although the analysis of interaction effects is beyond the purpose of the present article, it is worth discussing briefly what can be expected when the Hamiltonian (\ref{H0}) is replaced by a helical Luttinger liquid (HLL) Hamiltonian. In the first instance, because the HLL includes both intra- and inter-branch density-density interactions, the dynamical evolutions of the fields $\Psi_\pm$ would be no longer decoupled. As a consequence, while in the non-interacting case the photoexcited density profiles $\Delta n_{\pm}(x,t)$ re-acquire their character of left and right movers after the ending of the pulse, in the presence of interactions,  even after the pulse, each $\Delta n_{\pm}$ becomes a combination of both  right- and left-moving components, consisting of quasi-particles with a fractional charge ${\rm e}^*={\rm e} K$ travelling with a velocity $v=v_F/K$. 
Secondly, an important question is the dependence of the photoexcited observables on the temperature $T$ and chemical potential $\mu$. For the non-interacting case we have shown that the $T$- and $\mu$-independence of the photoexcited spatial profiles $\Delta_{\pm}(x,t)$ is related to the short-distance behavior of the correlations functions (\ref{correlations0}), determined by their scaling laws. Because interactions modify but do not destroy the scaling properties of the correlation functions, we expect the independence from $T$ and $\mu$ to be robust to interaction effects. In contrast, the photoexcited momentum distribution $\Delta f(k)$, which involves also long distance correlations, would be affected by interactions, and non analytical behavior are expected.
While interaction effects are often masked in dc measurements by the non interacting leads\cite{safi1995,maslov1995,ponomarenko1995}, they may become observable in time-resolved or finite-frequency measurements, as also emphasised in various works\cite{dolcini2005,dolcini2012,sassetti-dolcetto2015,kamata2014}.
We point out that the natural framework to describe interaction effects is the Bosonisation formalism\cite{vondelft} where, by expressing electron field operators $\Psi_\pm$ as exponential of bosonic fields $\Phi_\pm$, the interacting HLL Hamiltonian can be rewritten as a simply quadratic Hamiltonian for the latter fields. When the electric pulse is applied, $\Phi_\pm$ acquires a zero mode $\phi_\pm$, whose non-interacting limit precisely corresponds to the phases~$\phi_{\pm}$ given in Eqs.(\ref{phi+})-(\ref{phi-}). The spatial derivative of the zero mode, up to an additional term ensuring gauge invariance, is related to the photoexcited densities $\Delta n_{\pm}$, similarly to the third lines of Eqs.(\ref{Deltan_{+}_res-gen}) and (\ref{Deltan_{-}_res-gen}). 

{\it Possible implementation.}
Let us now discuss possible implementations of the proposed setup. QSH edge states have been observed in both HgTe/CdTe and in InAs/GaSb quantum wells, where they exhibit a linear dispersion with a Fermi velocity $v_F \simeq 5 \times 10^5 {\rm m/s}$ and $v_F \simeq 2 \times 10^4 {\rm m/s}$,  respectively\cite{molenkamp-zhang_jpsj,knez_2014}, within a bulk gap   $E_g \sim 30 \,{\rm meV}$. In these systems the phase breaking length $L_\phi$, i.e. the length scale for which the  analysis carried out here holds, is of the order of $1-2 \, \mu {\rm m}$ at Kelvin temperatures.
In order to generate a localised electric pulse, the most straightforward way might be to utilise a side finger electrode, contacted to the QSH bar and biased by a pulse voltage $V(t)$, similarly to what has been done in GaAs/AlGaAs 2DEG\cite{glattli2013}. In this case the spatial extension of the electric pulse is determined by the lateral width $\Delta \sim 100 {\rm nm}$ of the finger electrode, and the applied frequency range is in the GHz range. Alternatively, more localised electric pulses can be obtained with near field scanning optical microscopes (NSOM)\cite{novotny_review} in the illumination mode, as has been done both in semiconductor quantum dots\cite{koch1997}, in SWNTs\cite{novotny2003} and in QH chiral edge states\cite{nomura2011,nomura2015}: an optical fiber with a thin aperture of tens of ${\rm nm}$, positioned near the sample, excites a strong electric field at the tip apex due to an antenna effect. In this case the THz frequency regime is accessible. The recent impressive advances in pump-probe experiments and photo-current spectroscopy, successfully applied to time-resolved measurements in 2DEG\cite{glattli2013}, QH edge states\cite{bocquillon2014,lesueur2009}, graphene\cite{holleitner2014,jarillo-herrero2016} and also to the surface states of 3D topological insulators\cite{holleitner2015}, make the detection of the photoexcited wave packets in QSH system at experimental reach in the near future.

{\it Future developments.}  We conclude by outlining some possible developments of the present work within the field of electron quantum optics. We note that a quantum point contact (QPC) realised by etching a constriction in the quantum well could be used as a beam splitter on the electron wave packets photoexcited on one edge of the QSH bar, similarly to what is currently done for the edge states of QH systems\cite{bocquillon2014}. In the QSH case, however, due to the spin-orbit coupling characterising these materials, both spin-preserving and spin-flipping inter-edge tunneling terms may emerge across the constriction\cite{zhang-PRL,teo-kane2009,trauz-recher,dolcetto-sassetti2012,sassetti-ferraro2013,sternativo1,dolcini2015}. 
Due to the helical nature of the QSH states, the control of tunneling properties at the QPC might then open up the perspective to electrically  manipulate the spin of the photoexcited wave packets and their partitioning into various terminals\cite{chamon2009,richter,dolcini2011,citro-sassetti,dolcetto2014}. \\Another interesting development may be related to the observability of the so called `levitons'. These somewhat minimal quasiparticles, characterised by purely particle or purely hole excitations, were predicted by Levitov and coworkers\cite{levitov1996,levitov1997,levitov2006} to emerge as a response to Lorentzian-shaped voltage pulses of quantized area. After their recent observation in 1D channels created in ordinary semiconductor 2DEGs\cite{glattli2013}, they are on the spotlight in electron quantum optics\cite{flindt2015,moskalets2016,martin-sassetti_2016}, and it would thus be interesting to determine whether spin-polarised levitons can be generated in QSH edge channels. To this purpose, a time-domain analysis of the phases $e^{\pm i\phi_\pm}$ acquired by the two-counter-propagating electron fields (\ref{Psi-sol-gen}) is needed. The present work may provide the natural framework to address this problem: the general expressions (\ref{phi+}) and (\ref{phi-}) obtained in Sec.\ref{sec-3} hold for arbitrarily shaped pulses applied to the QSH edge states, and are not limited to the example of gaussian pulse discussed in Sec.\ref{sec-4}. Work is in progress along these lines.\\

\acknowledgments

Illuminating discussions with R. Rosati and E. Bocquillon are greatly acknowledged. F.D. also acknowledges financial support by the Italian FIRB 2012 project HybridNanoDev (Grant No.~RBFR1236VV).


\begin{thebibliography}{99}

\bibitem{bertoni2000} A. Bertoni, P. Bordone, R. Brunetti, C. Jacoboni, S. Reggiani, Phys. Rev. Lett. {\bf 84}, 5912 (2000).









\bibitem{feve2007} G. F\`eve,  A. Mah\'e, J.-M. Berroir,  T. Kontos,  B. Pla\c{c}ais,  D. C. Glattli, 
A. Cavanna,  B. Etienne,  Y. Jin, Science {\bf 316}, 1169 (2007).

\bibitem{feve2011} Ch. Grenier, R. Herv\'e, E. Bocquillon, F. D. Parmentier, B. Pla\c{c}ais, J. M. Berroir, G. F\`eve and P. Degiovanni, New J. Phys. {\bf 13}, 093007 (2011).


\bibitem{bocquillon2013} E. Bocquillon, V. Freulon, J.-M. Berroir, P. Degiovanni, B. Pla\c{c}ais, A. Cavanna, Y. Jin, and G. F\`eve, Nature Commun. {\bf 4}, 1839 (2013).

\bibitem{kataoka2013} J. D. Fletcher,  P. See,  H. Howe,  M. Pepper,  S. P. Giblin,  J. P. Griffiths,  G. A. C. Jones, I. Farrer,  D. A. Ritchie,  T. J. B. M. Janssen, and M. Kataoka, Phys. Rev. Lett. {\bf  111}, 216807 (2013).

\bibitem{bocquillon2014} E. Bocquillon, V. Freulon, F.D. Parmentier, J.-M Berroir, B. Pla\c{c}ais, C. Wahl, J. Rech, T. Jonckheere, T. Martin, C. Grenier, D. Ferraro, P. Degiovanni, G. F\`eve, Ann. Phys. (Berlin) {\bf 526}, 1 (2014).

\bibitem{feve2015} V. Freulon, A. Marguerite, J.-M. Berroir, B. Pla\c{c}ais, A. Cavanna, Y. Jin, and G. F\`eve,   Nature Commun. {\bf 6}, 6854 (2015).

\bibitem{kataoka2015} J. Waldie,  P. See,  V. Kashcheyevs,  J. P. Griffiths,  I. Farrer,  G. A. C. Jones,  D. A. Ritchie, T. J. B. M. Janssen,  and M. Kataoka, Phys. Rev. B {\bf 92}, 125305 (2015).


\bibitem{janssen2016} M. Kataoka,  N. Johnson,  C. Emary, P. See,  J. P. Griffiths,  G. A. C. Jones,  I. Farrer, D. A. Ritchie,  M. Pepper, and T. J. B. M. Janssen, Phys. Rev. Lett. {\bf 116}, 126803 (2016).

\bibitem{roulleau2008} P. Roulleau, F. Portier, P. Roche, A. Cavanna, G. Faini, U. Gennser, and D. Mailly, 
Phys. Rev. Lett. {\bf 100}, 126802 (2008).
\bibitem{kane-mele2005a} C. L. Kane, E. J. Mele, Phys. Rev. Lett. {\bf 95}, 146802 (2005).
\bibitem{kane-mele2005b} C. L. Kane, E. J. Mele, Phys. Rev. Lett. {\bf 95}, 226801 (2005).
\bibitem{bernevig_science_2006} B. A. Bernevig, T. L. Hughes, and S.-C. Zhang, Science {\bf 314}, 1757 (2006).

\bibitem{konig_2006}  M. K\"onig, S. Wiedmann, C. Br\"une, A. Roth, H. Buhmann, L. W. Molenkamp, X.-L.  Qi, and S.-C. Zhang, Science {\bf 318}, 766 (2006).
\bibitem{molenkamp-zhang_jpsj} M. K\"onig, H. Buhmann, L. W. Molenkamp, T. L. Hughes, C.-X. Liu, X.-L. Qi, and S.-C. Zhang, J. Phys. Soc. Jpn. {\bf 77}, 031007 (2008).
\bibitem{roth_2009} A. Roth, C.~Br\"une, H. Buhmann, L. W. Molenkamp, J.~ Maciejko, X.-L. Qi, and S.-C. Zhang, Science {\bf 325}, 294 (2009)
\bibitem{brune_2012} C. Br\"une,	A. Roth, H. Buhmann,	E. M. Hankiewicz, L. W. Molenkamp, J. Maciejko, X.-L. Qi, and S.-C. Zhang, Nature Phys. {\bf 8}, 485 (2012).
 
 
\bibitem{liu-zhang_2008} C. Liu, T. L. Hughes,  X.-L. Qi, K. Wang, and S.-C. Zhang, Phys. Rev. Lett. {\bf 100}, 236601 (2008).
\bibitem{knez_2007} I. Knez, R.-R. Du, and G. Sullivan, Phys. Rev. Lett. {\bf 107}, 136603 (2011).
\bibitem{knez_2014} I. Knez, C. T. Rettner, S.-H. Yang, and S. S. P. Parkin, L. Du,  R.-R. Du, and 
G. Sullivan, Phys. Rev. Lett. {\bf 112}, 026602 (2014).
\bibitem{spanton_2014} E. M. Spanton, K C. Nowack, L. Du, G. Sullivan,  R.-R. Du, and K. A. Moler, Phys. Rev. Lett. {\bf 113}, 026804 (2014).



\bibitem{rosati2015} R. Rosati, F.Dolcini and F. Rossi, Appl. Phys. Lett. {\bf 106}, 243101 (2015).


\bibitem{cayssol2012} B. D\'ora, J. Cayssol, F. Simon, and R. Moessner, Phys. Rev. Lett. {\bf 108}, 056602 (2012).
\bibitem{artemenko2013} S. N. Artemenko, and V. O. Kaladzhyan,  JETP Lett. {\bf 97}, 82 (2013). 
\bibitem{dolcetto-sassetti2014} G. Dolcetto, F. Cavaliere, and M. Sassetti, Phys. Rev. B {\bf  89}, 125419 (2014).
\bibitem{kindermann2009} M. J. Schmidt, E. G. Novik, M. Kindermann, and B. Trauzettel, Phys. Rev. B {\bf  79}, 241306(R) (2009).

\bibitem{artemenko2015} V. Kaladzhyan, P. P. Aseev,  and S. N. Artemenko, Phys. Rev. B {\bf 92}, 155424 (2015).



\bibitem{adler1969} S. L. Adler, Phys. Rev. {\bf 177}, 2426 (1969). 
\bibitem{bell-jackiw1969} J.S. Bell and R. Jackiw, Nuovo Cim. {\bf A 60}, 47 (1969).


\bibitem{burkov2012} A. A. Zyuzin and A. A. Burkov, Phys. Rev. B {\bf 86}, 115133 (2012).
\bibitem{li2013} H.-J. Kim,  K.-S. Kim, J.-F. Wang,  M. Sasaki,  N. Satoh, A. Ohnishi,  M. Kitaura,  M. Yang,  and L. Li, Phys. Rev. Lett {\bf 111}, 246603 (2013).
\bibitem{wishvanath2014} S. A. Parameswaran, T. Grover,  D. A. Abanin,D. A. Pesin,  and A. Vishwanath, Phys. Rev. X {\bf 4}, 031035 (2014)
\bibitem{takane2016} Y. Takane, J. Phys. Soc. Jpn {\bf 85}, 013706 (2016).
 

 


\bibitem{chen2014} Z. K. Liu, B. Zhou, Y. Zhang, Z. J. Wang, H. M. Weng,
D. Prabhakaran, S.-K. Mo, Z. X. Shen, Z. Fang, X. Dai, Z. Hussain, and Y. L. Chen, Science {\bf 343}, 864 (2014).
\bibitem{hasan2015a} S.-Y. Xu, I. Belopolski, N. Alidoust, M. Neupane, G.Bian, C. Zhang, R. Sankar, G. Chang, Z. Yuan, C.-C. Lee, S.-M. Huang, H. Zheng, J. Ma, D. S. Sanchez, B.Wang, A. Bansil, F. Chou, P. P. Shibayev, H. Lin, S. Jia, and M. Z. Hasan, Science {\bf 349}, 613 (2015).
\bibitem{hasan2015b} S.-Y. Xu, I. Belopolski, D. S. Sanchez, C. Zhang, G.
Chang, C. Guo, G. Bian, Z. Yuan, H. Lu, T.-R. Chang, P.
P. Shibayev, M. L. Prokopovych, N. Alidoust, H. Zheng,
C.-C. Lee, S.-M. Huang, R. Sankar, F. Chou, C.-H. Hsu,
H.-T. Jeng, A. Bansil, T. Neupert, V. N. Strocov, H. Lin,
S. Jia, and M. Z. Hasan, Science {\bf 349}, 622 (2015).
\bibitem{ong2015} J. Xiong, S. K. Kushwaha, T. Liang, J. W. Krizan, M.
Hirschberger, W. Wang, R. J. Cava, and N. P. Ong, Science
{\bf 350}, 413 (2015).

\bibitem{trauzettel2016} C. Fleckenstein, N. Traverso Ziani,  and B. Trauzettel,  arXiv:1607.05982.

\bibitem{bertlmann} R. A. Bertlmann, {\it Anomalies in quantum field theory}, (Clarendon Press, Oxford, 1996).

\bibitem{note-Dirac} Equation (\ref{eom-Psi}) is straightforwardly rewritten as $\gamma^\mu (i\hbar \partial_\mu -\frac{\rm e}{c} A_\mu) \Psi=0$ upon defining $\gamma^0=-\sigma_y$, $\gamma^1=-i \sigma_x$, $\left\{ A_\mu \right\} =(A_0,  A_1) = (\frac{c}{v_F} V,-A)$, and $\left\{ \partial_\mu \right\}= (\frac{1}{v_F}\partial_t \, , \,\partial_x)$.



\bibitem{vondelft} J. von Delft and H. Schoeller, Ann. Phys. (Leipzig) {\bf 7}, 225 (1998).
\bibitem{nielsen1983} H. B. Nielsen, N. Ninomiya, Phys. Lett. {\bf B130}, 389 (1983).



\bibitem{budich2012} J. C. Budich, F. Dolcini,  P. Recher,  and B. Trauzettel, Phys. Rev. Lett. {\bf 108}, 086602 (2012).
\bibitem{teo-kane2009}  J. C. Y. Teo and C. L. Kane, Phys. Rev. B {\bf 79}, 235321 (2009).

\bibitem{zhangPRL2009} J. Maciejko,  C. Liu, Y. Oreg, X.-L. Qi, C. Wu,  and S.-C. Zhang, Phys. Rev. Lett. {\bf 102}, 256803 (2009).
\bibitem{du2015} T. Li, P. Wang, H. Fu, L. Du, K. A. Schreiber, X. Mu, X. Liu, G. Sullivan, G. A. Cs\'{a}thy, X. Lin, R.-R. Du, Phys. Rev. Lett. {\bf 115}, 136804 (2015).



\bibitem{safi1995}
I. Safi and H.~J. Schulz, Phys. Rev. B {\bf 52},  R17040  (1995).

\bibitem{maslov1995}
D. L. Maslov and M. Stone, Phys. Rev. B {\bf 52},  R5539  (1995).

\bibitem{ponomarenko1995}
V.~V. Ponomarenko, Phys. Rev. B {\bf 52},  R8666  (1995).


\bibitem{dolcini2005} F. Dolcini, B. Trauzettel, I. Safi, and H.~Grabert, Phys. Rev. B {\bf 71}, 165309 (2005).
\bibitem{dolcini2012} F. Dolcini, Phys. Rev. B {\bf 85} 033306 (2012). 
\bibitem{sassetti-dolcetto2015} A. Calzona, M. Carrega, G. Dolcetto, and M. Sassetti, Phys. Rev. B {\bf  92}, 195414 (2015).


\bibitem{kamata2014} H. Kamata, N. Kumada, M. Hashisaka, K. Muraki, and T. Fujisawa, Nature Nanotech. {\bf 9}, 177 (2014).


\bibitem{glattli2013} J. Dubois, T. Jullien, F. Portier, P. Roche, A. Cavanna, Y. Jin, W. Wegscheider, P. Roulleau, and D. C. Glattli, Nature {\bf 502}, 659 (2013).

\bibitem{novotny_review} L. Novotny, and S. J. Stranick, Ann. Rev. Phys. Chem. {\bf 57}, 303 (2006).
\bibitem{koch1997} B. Hanewinkel, A. Knorr, P. Thomas, and S.W. Koch, Phys. Rev. B {\bf 55}, 13715 (1997).
\bibitem{novotny2003} A. Hartschuh, E. J. S\'anchez,  X. S. Xie,  and L. Novotny, Phys. Rev. Lett. {\bf 90}, 095503 (2003).
\bibitem{nomura2011} H. Ito,  K. Furuya,  Y. Shibata,  S. Kashiwaya,  M. Yamaguchi,  T. Akazaki,  H. Tamura,  Y. Ootuka,  and S. Nomura, Phys. Rev. Lett. {\bf 107}, 256803 (2011).
\bibitem{nomura2015} S. Mamyouda,  H. Ito, Y. Shibata, S. Kashiwaya, M. Yamaguchi, T. Akazaki, H. Tamura, Y. Ootuka, and S. Nomura, Nanolett. {\bf 15}, 2417 (2015).


\bibitem{lesueur2009} C. Altimiras, H. le Sueur, U. Gennser, A. Cavanna, D. Mailly and F. Pierre, Nature Phys. {\bf 6}, 34 (2009).


\bibitem{holleitner2014} A. Brenneis, L. Gaudreau, M. Seifert, H. Karl, M. S. Brandt, H. Huebl,
J. A. Garrido, F. H. L. Koppens, and A. W. Holleitner, Nature Nanotech. {\bf 10}, 135 (2014).
\bibitem{jarillo-herrero2016} A. Woessner, P. Alonso-Gonz\'alez, M. B. Lundeberg, Y. Gao, J. E. Barrios-Vargas,
G. Navickaite, Q. Ma, D. Janner, K. Watanabe, Aron W. Cummings, T. Taniguchi,
V. Pruneri, S. Roche, P. Jarillo-Herrero6, James Hone4, Rainer Hillenbrand, and F. H. L. Koppens, Nature Commun. {\bf 7}, 10783 (2016).
\bibitem{holleitner2015} C. Kastl, C. Karnetzky, H. Karl, and A. W. Holleitner, Nature Commun. {\bf 6}, 6617 (2015).





 

\bibitem{zhang-PRL} C. Wu, B. A. Bernevig, and S-C. Zhang, Phys. Rev. Lett. {\bf 96}, 106401 (2006).

\bibitem{trauz-recher} C.-X. Liu, J.~C.~Budich, P.~Recher, and B.~Trauzettel,  Phys. Rev. B {\bf 83}, 035407 (2011).
\bibitem{dolcetto-sassetti2012} G. Dolcetto, S. Barbarino, D. Ferraro, N. Magnoli, and M. Sassetti, Phys. Rev. B {\bf 85}, 195138 (2012).
\bibitem{sassetti-ferraro2013} D. Ferraro,  G. Dolcetto,  R. Citro, F. Romeo,  and M. Sassetti, Phys. Rev. B {\bf  87}, 245419 (2013).
\bibitem{sternativo1} P. Sternativo, and F. Dolcini, Phys. Rev. B {\bf 89}, 035415 (2014).
\bibitem{dolcini2015} F. Dolcini, Phys. Rev. B {\bf 92}, 155421 (2015).
\bibitem{chamon2009} C.-Y. Hou,  E.-A. Kim, and C. Chamon, Phys. Rev. Lett. {\bf  102}, 076602 (2009).
\bibitem{richter} V. Krueckl and K. Richter, Phys. Rev. Lett. {\bf 107}, 086803 (2011).
\bibitem{dolcini2011} F. Dolcini, Phys. Rev. B {\bf 83} 165304 (2011).
\bibitem{citro-sassetti} F. Romeo, R. Citro, D. Ferraro, and M. Sassetti, Phys. Rev. B {\bf 86}, 165418 (2012).
\bibitem{dolcetto2014} G. Dolcetto, L. Vannucci,  A. Braggio,  R. Raimondi,  and M. Sassetti, Phys. Rev. B {\bf 90}, 165401 (2014).


\bibitem{levitov1996} L. S. Levitov, H. Lee, and G. B. Lesovik, J. Math. Phys. {\bf 37}, 4845 (1996).
\bibitem{levitov1997} D. A. Ivanov, H.W. Lee, and L. S. Levitov, Phys. Rev. B {\bf 56}, 6839 (1997).
\bibitem{levitov2006} J. Keeling, I. Klich, and L. S. Levitov, Phys. Rev. Lett. {\bf 97}, 116403 (2006).
\bibitem{flindt2015} D. Dasenbrook, and C. Flindt, Phys. Rev. B {\bf 92}, 161412 (2015).
\bibitem{moskalets2016} M. Moskalets, Phys. Rev. Lett. {\bf 117}, 046801 (2016).
\bibitem{martin-sassetti_2016} J. Rech, D. Ferraro, T. Jonckheere, L. Vannucci, M. Sassetti, and T. Martin,   arXiv:1606.01122.
\end{thebibliography}
\end{document}